\documentclass[11pt]{article}
\usepackage{geometry}                
\usepackage{graphicx}
\usepackage{amssymb}
\usepackage{epstopdf}
\usepackage{color}
\usepackage{soul}
\usepackage[shortlabels]{enumitem}
\usepackage{multirow}
\usepackage{longtable}
\usepackage{subfigure}

\newcommand{\bs}{\mathbf{s}}

\newcommand{\bz}{\mathbf{z}}
\newcommand{\bx}{\mathbf{x}}

\newcommand{\E}[1]{\mathbb{E}\left[{#1}\right]}

\newcommand{\EQ}[1]{\mathbb{E}_{Q}\left[{#1}\right]}
\newcommand{\Lik}{\mathcal{L}}

\begin{document}


\title{A simple probabilistic neural network for machine understanding}

\author{%
  Rongrong Xie \\
  \small{Key Laboratory of Quark and Lepton Physics (MOE) and Institute of Particle Physics,} \\
  \small{Central China Normal University (CCNU), Wuhan, China} \\
  and \\
  Matteo~Marsili\thanks{marsili@ictp.it} \\
    \small{Quantitative Life Sciences Section}\\
    \small{The Abdus Salam International Centre for Theoretical Physics,
  34151 Trieste, Italy}}

\date{}

\maketitle


\begin{abstract}
We discuss probabilistic neural networks with a fixed internal representation as models for machine understanding. Here understanding is intended as mapping data to an already existing representation which encodes an {\em a priori} organisation of the feature space. 

We derive the internal representation by requiring that it satisfies the principles of maximal relevance and of maximal ignorance about how different features are combined. We show that, when hidden units are binary variables, these two principles identify a unique model -- the Hierarchical Feature Model (HFM) -- which is fully solvable and provides a natural interpretation in terms of features. 

We argue that learning machines with this architecture enjoy a number of interesting properties, like the continuity of the representation with respect to changes in parameters and data, the possibility to control the level of compression and the 
ability to support functions that go beyond generalisation.

We explore the behaviour of the model with extensive numerical experiments and argue that models where the internal representation is fixed reproduce a learning modality which is qualitatively different from that of  traditional models such as Restricted Boltzmann Machines. 
\end{abstract}


\begin{quote}
"What I cannot create, I do not understand"\\ 
\rightline{(Richard P. Feynman)}
\end{quote}


The advent of machine learning has expanded our ability to ``create'', i.e. to generalise from a set of examples, much beyond what we understand. 
In the classical view, the ability to create data entails extracting compressed representations of complex data that capture their regularities. In this view, the simpler the algorithm~\cite{chaitin2006limits} or the statistical model~\cite{myung} that captures the regularities of the data, the more we understand. Yet, even machines like auto-encoders or deep belief networks extract compressed representations from complex data. But theirs is a form of understanding that is unintelligible to us. 

Furthermore, the many triumphs of machine intelligence from automatic translation to the generation of texts and images, have shown that the ability to ``create'' does not require simplicity. The accuracy in deep neural networks can increase with complexity (i.e. with the number of parameters), without overfitting~\cite{zhang2021understanding,mei2019generalization}. 

In this paper "understanding" will be interpreted as "representing new knowledge or data within a preexisting framework or model\footnote{We thank Alessandro Ingrosso for suggesting this interpretation of ``understanding''.}". Our aim is to explore the properties of a learning modality, in which the internal representation is fixed, within an unsupervised learning framework of simple one layer probabilistic neural networks. 
When the internal representation is fixed {\em a priori}, only the conditional distribution of the data, given the hidden variables, needs to be learned. 

The models we shall study are similar in spirit to variational auto-encoders~\cite{kingmaauto,bengio2013representation}. Yet they differ in how the internal representation is chosen. Rather than insisting on computational manageability or on independence of the hidden features~\cite{bengio2013representation,locatello2019challenging}, we derive the distribution of hidden variables from first principles. We do this by drawing from previous work~\cite{marsili2022quantifying} that introduces an absolute notion of {\em relevance} for representations and argues that relevance should be maximal for internal representations of neural networks trained on datasets with a rich statistical structure. 
Indeed, the {\em principle of maximal relevance}~\cite{marsili2022quantifying} has been shown to characterise both trained models in machine learning~\cite{Song,Odilon} and efficient coding~\cite{cubero2018minimum} (see Section~\ref{sec:maxrel} for more details). For models with binary hidden variables, which are those we focus on, we show that the principle of maximal relevance and the requirement that the occurrence of features be as unbiased as possible, uniquely identify a model. This model exhibits a hierarchical organisation of the feature space, which is why we call it the {\em Hierarchical Feature Model} (HFM).

There are several reasons why such an approach may be desirable, besides being supported by first principles: 
it allows the internal representation to be chosen as a simple model, in agreement with the Occam razor, thus providing a transparent interpretation of the features. Models where the internal representation is fixed operate in a statistical regime characterised by the classical bias--variance tradeoff, which is qualitatively different from the regime in which over-parametrised (probabilistic) neural networks, such as Restricted Boltzmann Machines~\cite{hinton2012practical,decelle2021restricted} (RBMs), operate. There are many more properties that differ markedly from those that hold in RBMs, that we shall take as the natural reference point for our model:
The compression level of the internal representation can be fixed at the outset. There is no need to resort to stochastic gradient descent based training. Features can be introduced one by one in the training protocol and, as we shall see, already learned features need not be learned anew when new features are introduced. 
These properties do not hold in RBMs\footnote{We refer to classical RBM defined in Appendix~\ref{app:RBM}. C{\^o}t{\'e} and Larochelle~\cite{cote2016infinite} introduce a version of the RBM where hidden units are hierarchically ordered and in which features can be learned one by one, and their number does not need to be set {\em a priori}.}.
Also, sampling the equilibrium distribution of data is trivial when the internal representation is chosen {\em a priori}, while it is not in RBMs~\cite{decelle2021equilibrium}. 
Finally, a fixed internal representation makes it possible to explore functions that go beyond learning and generalisation, as we shall discuss later. 



After illustrating these points on few case studies, we will argue that learning, in our model, follows a qualitatively different modality with respect to the one of RBMs. In RBMs, data specific information is transferred to the internal representation, with weights which encode generic features~\cite{decelle2021restricted}.  
In contrast, when the internal representation is fixed, data specific information is necessarily incorporated in the weights. Hence weights can be thought of as  feature vectors extracted from the data. In a design where the internal representation is fixed, training is akin to learning how to organise the data into an abstract predetermined archive. We speculate that the abstraction implicit in this training modality may support higher ``cognitive'' functions, such as making up correlations between data learned separately. We shall comment further on the difference between these two learning modalities in the concluding Section, besides discussing further research avenues.


\section{Single layer probabilistic neural networks}
\label{sec:one}

We focus on unsupervised learning and, in particular, on probabilistic single layer neural networks composed by a vector of $m$ binary variables $\bx=(x_1,\ldots,x_m)$ coupled to a vector $\bs=(s_1,\ldots,s_n)$ of $n$ binary variables ($x_j,s_i\in\{0,1\}$). $\bx$ is called the visible layer whereas $\bs$ is the hidden layer. The network is defined by a joint probability distribution $p(\bx,\bs|\theta)$ over the visible and the hidden variables, that depends on a vector $\theta$ of parameters.

Given a dataset $\hat \bx=(\bx^{(1)},\ldots,\bx^{(N)})$ of $N$ samples of the visible variables, the network is trained by maximising the log-likelihood
\begin{equation}
\label{ }
\mathcal{L}(\hat\bx|\theta)=\sum_{i=1}^N \log p(\bx^{(i)}|\theta),
\end{equation}
over $\theta$, where
\begin{equation}
\label{ }
p(\bx|\theta)=\sum_{\bs} p(\bx,\bs|\theta)
\end{equation}
is the marginal distribution of $\bx$. We'll denote by $\hat \theta={\rm arg}\max_\theta \mathcal{L}(\hat\bx|\theta)$ the learned parameters (after training). Training maps the dataset $\hat \bx$ onto an internal representation 
\begin{equation}
\label{eq:ps}
p(\bs|\hat\theta)=\sum_{\bx} p(\bx,\bs|\hat\theta)
\end{equation}
which is the marginal distribution over hidden states $\bs$.

For example, in a RBM (see Appendix \ref{app:RBM}) the joint distribution $p(\bx,\bs|\theta)$ is defined in such a way that the variables $\bx$ and $\bs$ are easy to sample by Markov chain Montecarlo methods. This allows one to estimate the gradients of the log-likelihood. During training of a RBM both the internal representation $p(\bs|\theta)$ and the conditional distribution $p(\bx|\bs,\theta)$ are learned from scratch. 

%
%
This paper explores architectures 
\begin{equation}
\label{brain}
p(\bx,\bs|\theta)=p(\bx|\bs,\theta)p(\bs)
\end{equation}
where the internal representation $p(\bs)$ is fixed at the outset, and only the conditional distribution $p(\bx|\bs,\theta)$ is learned. 

We assume that all statistical dependencies between the $x_j$ are ``explained'' by the variable $\bs$. This implies that, conditional on $\bs$, all components of $\bx$ are independent. For binary variables $x_j\in\{0,1\}$ this implies\footnote{A straightforward generalisation to continuous variables is possible, taking $x_j$ as Gaussian variables with mean $a_j+\sum_is_iw_{i,j}$ and variance $\sigma^2_j$.}
\begin{equation}
\label{ }
p(\bx|\bs,\theta)=\prod_{j=1}^m \frac{e^{h_j(\bs)x_j}}{1+e^{h_j(\bs)}},\qquad h_j(\bs)=a_j+\sum_{i=1}^n s_i w_{i,j}\,. 
\end{equation}
We shall interpret the vector $\vec{w}_i=(w_{i,1},\ldots,w_{i,m})$ as {\em feature} $i$. Hence points $\bx$ drawn from $p(\bx|\bs,\theta)$ with $s_i=1$ are characterised by feature $i$, whereas if $s_i=0$ the distribution $p(\bx|\bs,\theta)$ generates points without feature $i$. The internal configuration $\bs$ then specifies a profile of features. The distribution $p(\bs)$ encodes the way in which the space of features is populated.

Architectures where the internal representation is fixed at the outset and only the output layer is learned have been proposed for supervised learning tasks (see e.g.~\cite{rahimi2007random,kasun2013representational,principe2015universal}) and for unsupervised learning, e.g. in auto-encoders~\cite{bengio2013representation,kingmaauto}. Their success suggests that the internal representation may be largely independent of the data being learned.
The choice of $p(\bs)$ in these examples is dictated mostly by computational efficiency and/or by requirements of interpretability\footnote{Variational auto-encoders generally assume an internal representation where $\bs$ are independent Gaussian variables with unit variance. This enforces a representation of ``disentangled'' features, i.e. of features which are independent of each other. Locatello {\em et al.}~\cite{locatello2019challenging} discuss the limits of this approach.}.
Our focus is not on computational efficiency but rather on deriving $p(\bs)$ from first principles rooted in information theory. 

\subsection{The principle of maximal relevance}
\label{sec:maxrel}

The first requirement that we shall impose on $p(\bs)$ is that it should obey a principle of maximal relevance. Relevance has been recently proposed~\cite{marsili2022quantifying,MMR} as a context and model free measure of informativeness of a dataset or of a representation. It is defined as the entropy of the distribution of coding costs $E_{\bs}\equiv -\log_2 p(\bs)$, i.e. 
\begin{equation}
\label{ }
H[E]=-\sum_E p(E)\log_2 p(E),\qquad p(E)=\sum_{\bs}p(\bs)\delta\left(E+\log_2 p(\bs)\right)\,.
\end{equation}
The relevance differs from the Shannon entropy of $p(\bs)$, 
\begin{equation}
\label{eqHs}
 H[\bs]=-\sum_{\bs}p(\bs)\log_2 p(\bs)\equiv 
 \E{E_{\bs}} \,,
\end{equation}
which is the average coding cost\footnote{We follow the standard notation~\cite{CoverThomas} for the entropy $H[X]$ of a random variable $X$.}. $H[\bs]$ quantifies the compression level and, following Ref.~\cite{marsili2022quantifying}, it  is called {\em resolution}. 

Ref.~\cite{marsili2022quantifying} provides several arguments that support the hypothesis that efficient representations satisfy the principle of maximal relevance, i.e. that $p(\bs)$ maximises $H[E]$ at  a given average coding cost $H[\bs]$. 
For example, it shows that $H[E]$ lower bounds the mutual information between the hidden state $\bs$ of the network and the hidden features that the training process extracts from the data. So representations of maximal relevance are those that, in theory, extract the largest amount of information from data. 

The principle of maximal relevance dictates that the number $W(E)$ of states $\bs$ with coding cost $E=-\log_2 p(\bs)$ should follow an exponential law, $W(E)=W_0e^{g E}$, where the constant $g$ depends on the resolution $H[\bs]$. The exponential law entails an efficient use of information resources. In order to see this remember that the coding cost $E$ measures the level of detail of a state, i.e. the (minimal) number of bits needed to represent it. There are $2^E$ possible states with a level of detail $E$, hence $W(E)\propto 2^E$ corresponds to a situation in which the feature space is exploited optimally, in the sense that it is occupied as uniformly as possible\footnote{Notice that $\log_2 W(E)$ is the number of bits needed to distinguish between states with the same level of detail $E$. Hence $W(E)\propto 2^E$ implies that the information cost (in bits) of retrieving a state is the same, apart from an offset, as the number of bits needed to describe it, which is the description length $E$ (in bits) of that state.}. For a generic value of $H[\bs]$, the same principle leads to $W(E)=W_0e^{g E}$, as shown in~\cite{marsili2022quantifying}, with $g\neq \log 2$ which depends on the resolution $H[\bs]$.

As a function of $H[\bs]$, the maximal value of the relevance $H[E]$ has a non-monotonic behaviour which distinguishes two learning regimes. For large values of $H[\bs]$, $H[E]$ is a decreasing function of $H[\bs]$. In other words, compression, i.e. a reduction of $H[\bs]$, brings an increase in relevance in this regime. Learning in this regime is akin to compressing out irrelevant details (i.e. noise) from data. Upon decreasing $H[\bs]$ further, $H[E]$ reaches a maximum and then starts decreasing. The maximum of $H[E]$ corresponds to the most compressed lossless representation. This point coincides with the optimum discussed above where $W(E)\propto 2^E$. A further reduction of $H[\bs]$ beyond this point leads to lossy compressed representations. 

We refer to Ref.~\cite{marsili2022quantifying} for a broader discussion of the principle of maximal relevance. Let it suffice to say that Refs.~\cite{Song,Odilon} show that, within the limits imposed by the data and the architecture, the internal representation of  learning machines trained on a dataset with a rich structure obeys the principle of maximal relevance\footnote{When applied to training of a model from a dataset of $N$ samples, the principle of maximal relevance suggests that the population of the internal representation $p(\bs)$ should obey Zipf's law~\cite{marsili2022quantifying}. This implies that the number of sampled states $\bs$ should be proportional to $\sqrt{N}$. This should be smaller than the number $2^n$ of possible states, which leads us to a lower bound $n>\frac 1 2 \log_2 N$ for the number of hidden nodes. In the case of the MNIST dataset ($N=6\cdot 10^4$) which we shall discuss later, this yields $n>8$.}. 

\subsection{The Hierarchical Feature model}

We shall then choose an internal representation $p(\bs)$ that encodes an {\em a-priori} organisation of the feature space which is consistent with the principle of maximal relevance. In addition, our model assumes a hierarchical space of features in which the index $k$ of $s_k$ denotes the level of detail of that feature. We require that $p(\bs)$ satisfies a principle of maximal {\em a priori} ignorance in the sense that conditional to $s_k=1$, all states $\bs=(s_1,\ldots,s_{k-1},1,0,\ldots,0)$ with $s_i=0,1$ for $i<k$ are equally likely. In other words, conditional to $s_k=1$, all $s_i$ for $i<k$ are independent and identically distributed with  $P\{s_i=1|s_k=1\}=\frac 1 2$.

As a direct consequence of this requirement, $p(\bs)$ must be a function of the largest index $i$ for which $s_i=1$, i.e. it must be a function of the level of detail\footnote{The state $\bs_0=(0,\ldots,0)$ has $m_{\bs}=0$.} 
\begin{equation}
\label{ }
m_\bs\equiv\max\{i:~s_i=1\}
\end{equation}
of state $\bs$. Since the number of states with $m_\bs=k$ equals $2^{k-1}$, the exponential behaviour of $W(E)$ dictated by the principle of maximal relevance implies that
\begin{equation}
\label{HFM}
p(\bs)=\frac 1 Z e^{-g\mathcal{H}(\bs)},\qquad \mathcal{H}(\bs)=\max\{m_\bs-1,0\}\,,
\end{equation}
where the partition function
\begin{equation}
\label{eqZ}
Z=\sum_{\bs}e^{-g\mathcal{H}(\bs)}=1+\frac{\xi^n-1}{\xi-1},\qquad \xi=2e^{-g}
\end{equation}
ensures normalisation. 

Notice that, {\em a priori}, the points with just the first feature ($s_1=1$ and $s_i=0$ for all $i>1$) and the one without any feature ($s_i=0$ for all $i$) are equiprobable. We call the distribution of Eq.~(\ref{HFM}) the {\em Hierarchical Feature Model} (HFM).

A system with an Hamiltonian given by Eq.~(\ref{HFM}) has a non-trivial statistical mechanics. In the micro-canonical ensemble, the distribution of $\bs$ with constant energy has fixed $m_{\bs}=k$, which means that all variables $s_i$ with $i<m_{\bs}$ are in a maximum entropy state -- as if at infinite temperature -- whereas all variables with $i\ge k$ are frozen -- as in a zero entropy state.

The expected value of the Hamiltonian $\mathcal{H}$ is easily computed, as well as the entropy (i.e. the resolution)
\begin{eqnarray}
\E{\mathcal{H}} & = & \xi\left[\frac{n\xi^{n-1}-1}{\xi^n+\xi-2}-\frac{1}{\xi-1}\right] \\
H[\bs] & = & \frac{g}{\log 2}\E{\mathcal{H}} +\log_2\left(1+\frac{\xi^n-1}{\xi-1}\right)\,.
\end{eqnarray}
We take $\E{\mathcal{H}}$ as an effective number of features used by the model.

\begin{figure}
  \centering
\includegraphics[width=0.8\textwidth,angle=0]{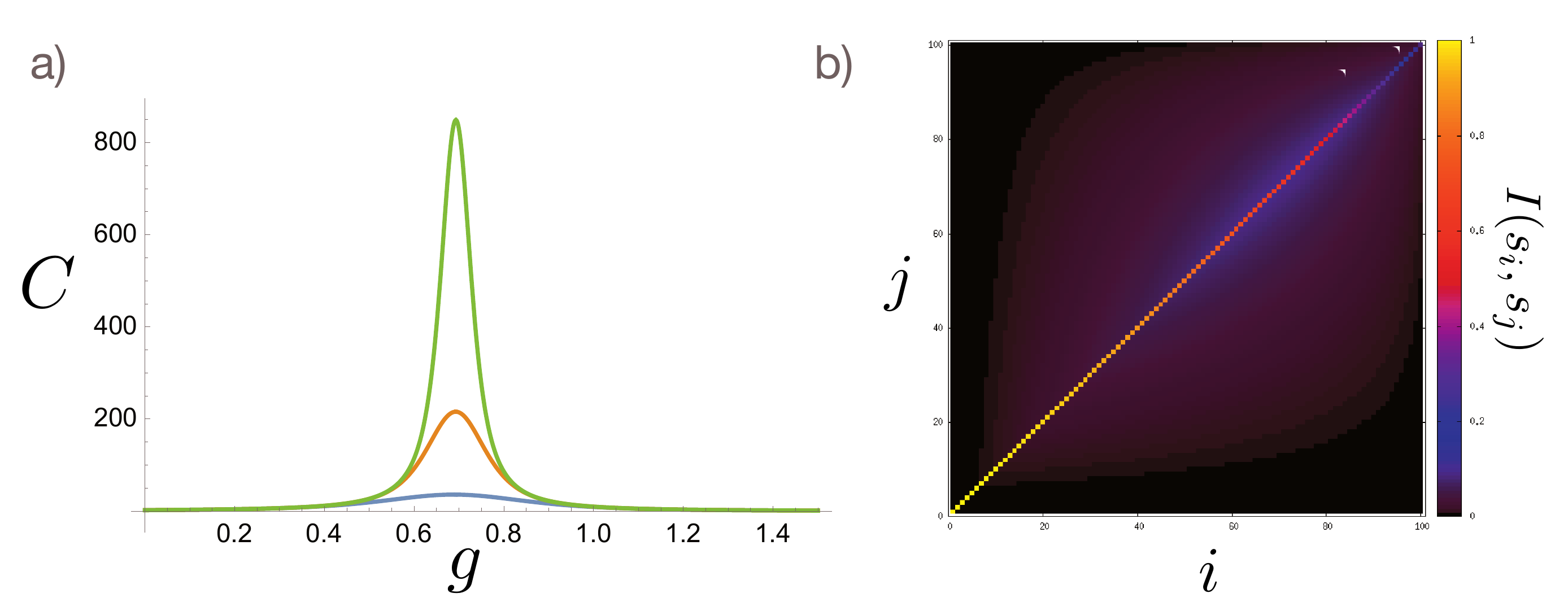}
  \caption{\label{fig:model} {\em a)} Specific heat of the HFM as a function of $g$ for $n=20$ (blue), $50$ (orange) and $100$ (green). {\em b)} Mutual information $I(s_i,s_j)$ (in bits) between different variables of the HFM, for $n=100$ and $g=g_c$. The value on the diagonal is the entropy $H[s_i]$ of variable $s_i$.}
\end{figure}

In the limit $n\to\infty$, the model exhibits a phase transition at $g_c=\log 2$ (where $\xi=1$). In the weak coupling phase ($g<g_c$) the model spans an extensive number of features and both the Hamiltonian and the entropy are extensive  ($\E{\mathcal{H}}, H[\bs]\sim n$). In the strong coupling phase ($g>g_c$) the typical number of features is finite and both $\E{\mathcal{H}}$ and $H[\bs]$ are of order one. 
Close to $g_c$ we find 
\[
\E{\mathcal{H}}\simeq \frac n 2 -1 +\frac{1}{n+1}+O\left(n^2(g-g_c)\right)
\]
which implies that for large $n$ the crossover between the two regimes occurs in a region of size
$|g-g_c|\sim 1/n$ around $g_c$. The phase transition is signalled by a divergence of the ``specific heat''\footnote{In statistics, $C(g)$ coincides with the Fisher Information. The integral of $\sqrt{C(g)}$ over $g\in[0,\infty)$ defines the stochastic complexity of the model $\mathcal{J}=\log\int_0^\infty\! dg\sqrt{C(g)}$. The numerical estimates for $n\le 500$ suggest that, for large $n$, the leading behaviour is given by $\mathcal{J}\simeq \log\log n+ {\rm const.}$.}
$C(g)=\E{\left(\mathcal{H}-\E{\mathcal{H}}\right)^2}$ at $g_c$, as shown in Fig.~\ref{fig:model}a). At $g_c$ we find
\begin{equation}
\label{ }
C(g_c)=\frac{n(n-1)[n(n+5)-2]}{12 (n+1)^2}
\simeq \frac{n^2}{12},\qquad (n\gg 1)
\end{equation}
when $n\to\infty$. This behaviour is consistent with a uniform distribution of $\mathcal{H}/n$ in its allowed range $[0,1]$, i.e. with the principle of maximal relevance~\cite{Odilon} that dictates that the distribution of coding costs should be as broad as possible. 

It is easy to compute moments of arbitrary order $\E{s_{i}s_j\cdots s_{\ell}}$. We refer the interested reader to Appendix~\ref{app:HFM} for a derivation. There we also show that the model $p(\bs)$ contains interactions of arbitrary order, because it can be written as an exponential model in terms of product operators $\phi_\nu(\bs)$ with interactions of arbitrary order, i.e. 
\begin{equation}
\label{ }
p(\bs)=\frac 1 Z e^{\sum_\nu J_\nu \phi_\nu(\bs)},\qquad \phi_\nu(\bs)=\prod_{i\in\nu} s_i\,.
\end{equation}
Appendix~\ref{app:HFM} shows that the couplings are given by the surprisingly simple formula
\begin{equation}
\label{Jnu_fin}
J_\nu=g(-1)^{|\nu|}(i_\nu-1)
\end{equation}
where $|\nu|$ is the order of the interaction (i.e. the number of variables in $\phi_\nu$) and $i_\nu$ is the smallest index of the variables $s_i$ that are present in the operator $\phi_\nu$. Note that $i_\nu=1$ for all operators that contain $s_1$,  and hence $J_\nu=0$. Indeed the first feature is independent of all the others, i.e. $p(\bs)=p(s_1)p(s_2,\ldots, s_n)$, which is evident from the fact that the Hamiltonian $\mathcal{H}(\bs)$ does not depend on $s_1$. Even though the higher order features $s_2,\ldots, s_n$ are not independent, Fig.~\ref{fig:model}b) shows that, even at the critical point, the mutual information $I(s_i,s_j)$ between features $i$ and $j$ is generally small, showing a high level of disentanglement of features. This is surprising in view of the fact that interactions of all order are present and that they are not small.

\section{Training learning machines with fixed $p(\bs)$}
\label{sec:occam}

Let us now consider the problem of training a model of the type of Eq.~(\ref{brain}). The log-likelihood can be written as 
\begin{equation}
\label{llik}
{\Lik}=\frac 1 N \log p(\hat \bx|\theta)=\overline{\log p(\bx|\theta)}=\overline{\log\E{p(\bx|\bs,\theta)}}
\end{equation}
where $\overline{g(\bx)}=\frac 1 N\sum_a g\left(\bx^{(a)}\right)$ denotes the average over the data of a generic function $g(\bx)$, and $\E{\ldots}$ stands for the expected value over $p(\bs)$. 

We shall study learning as features are added one by one. This is motivated by the Occam razor, that, in the present case, suggests that features should not be introduced unless they are needed\footnote{\label{foot:thermo} A further reason for studying algorithms where the internal state space expands gradually is given by thermodynamic efficiency~\cite{parrondo2015thermodynamics,still2020thermodynamic}. In brief, training brings the machine from an equilibrium state $p(\bx,\bs|\theta_0)$ to a final state $p(\bx,\bs|\hat\theta)$. The free energy difference 
can be written as $\Delta F=\Delta U-\Delta H[\bx,\bs]=W+Q-\Delta H[\bx,\bs]$. Here we used the first law of thermodynamics which states that the change $\Delta U$ in the energy is the sum of the work $W$ done on the system and the heat $Q$ absorbed by the system. The second law of thermodynamics, states that the work $W$ needed in the transformation is at least equal to $\Delta F$. Therefore $W-\Delta F=-Q+\Delta H[\bx,\bs]\ge 0$, i.e. the heat released to the environment during training is bounded by the change in the joint entropy $H[\bx,\bs]$, i.e.
\begin{equation}
\label{ }
-Q\ge-\Delta H[\bx,\bs]=I(\bx,\bs)-\Delta H[\bx]-\Delta H[\bs].
\end{equation}
Here $I(\bx,\bs)$ is the mutual information between the visible and the hidden layer after training, which is a measure of how much the machine has learnt. Other things being equal, the released heat is minimal in training protocols where $H[\bs]$ is initially small and it increases during training, i.e. when the state space gradually expands during training.
}. 

First we discuss some generic properties and then turn to numerical algorithms that we will deploy in the sequel on specific datasets.

\subsection{The featureless state}

Let us consider the {\em featureless state} $p(\bs)=\delta_{\bs,\bs_0}$ where $\bs_0=(0,0,\ldots,0)$ is the state where $s_i=0$ for all $i=1,\ldots,n$. This state corresponds to zero coding cost, i.e.  $H[\bs]=0$ and it is akin to a zero temperature state.
The probability $p(\bx|\bs,\theta)$ then depends only on the parameters $a_j$, and the likelihood is maximal when 
\begin{equation}
\label{w0sol}
a_j=\log\frac{\bar x_j}{1-\bar x_j},\qquad \forall j\,.
\end{equation}
In this state, the data does not provide any information on the features. With a flat prior on $\vec w_i$ the posterior remains a flat, totally uninformative distribution\footnote{It is interesting to note that, by Jensen's inequality, $\Lik\ge \overline{\E{\log p(\bx|s)}}$. The lower bound $\overline{\E{\log p(\bx|s)}}$ only depends on the data through the first moment $\bar x_j$ and it is maximal when $a_j$ is given by Eq.(\ref{w0sol}) and $w_{i,j}=0$ for all $i,j$.}. It is conceptually appealing that the featureless state corresponds to a state of maximal ignorance about the features.

\subsection{Learning one feature at a time}

With increasing resolution $H[\bs]$ we're led to study distributions where $p(\bs)$ extends to states $\bs\neq \bs_0$. Let us start with the state where $p(\bs)>0$ only for states $\bs_0$ and $\bs_1=(1,0,\ldots,0)$, which represents points with the first feature. This distribution corresponds to the limit $g\to\infty$ of the HFM.

Expanding around the state $w_{i,j}=0$ for all $i,j$, we find (see Appendix~\ref{app:w1})
\begin{equation}
\label{expw1text}
\Lik=\sum_j \left[\bar x_j\log\bar x_j+(1-\bar x_j)\log(1-\bar x_j)\right]+\frac 1 2 
\sum_{j,j'}w_{1,j}D_{j,j'}w_{1,j'} + O(w^3)
\end{equation}
with
\[
D_{j,j'} =  \frac 1 4 
\left[\overline{(x_j-\bar x_j)(x_{j'}-\bar x_{j'})}-\bar x_j(1-\bar x_j)\delta_{j,j'}\right]\,.
\]
Inspection of the matrix $\hat D$ reveals\footnote{Notice that the diagonal elements of $\hat D$ vanish, hence its trace also vanishes. This implies that there are positive and negative eigenvalues.} that the state $\hat w=0$ is a saddle point. This means that second order statistics of the data provides information on the orientation where the likelihood increases but not on whether it increases faster in the positive or the negative direction. This information is contained in $O(w^3)$ terms that depend on higher order moments (see Appendix~\ref{app:w1}). 
This is a further manifestation of the fact that relevant statistical information in structured data is carried by higher order moments (see e.g. \cite{ingrosso2022data}). 

The gradients of the log-likelihood with respect to the second feature $w_{2,j}$ also vanish when $w_{2,j}=0$, because of the independence between $s_1$ and $s_2$. The gradient ascent dynamics needs to start from a non-zero value of $w_{i,j}$ for $i=1,2$. The same is true for RBMs and the standard recipe to overcome this difficulty is to start training from a randomly chosen value of $w_{i,j}$~\cite{hinton2012practical,decelle}. 

In our case, once the first two features are learned, the next features can be learned starting from the $\vec w_i=0$ state, because the gradients of the log-likelihood are non-zero at this point (we refer to Appendix~\ref{app:w1} for a detailed analysis). 

\subsection{An expectation maximisation algorithm}

In spite of the simplicity of the HFM, the calculation of the likelihood $\log p(\hat\bx|\theta)$ or of its gradients is a complex task. It is in principle possible to devise Markov chain Montecarlo algorithms to sample the joint distribution $p(\bx,\bs|\theta)$ or the conditional one $p(\bs|\bx,\theta)$, in order to estimate the gradients. In addition, as mentioned above, training the first two features requires starting the dynamics from a nonzero random initial condition for the first two features. 

We explore, instead, deterministic algorithms that maximise the joint log-likelihood $\overline{\log p(\bx,\bs|\theta)}$, exploiting the fact that samples from $p(\bs)$ can be drawn independently of the data\footnote{The reason for exploring deterministic algorithms is that, given that the internal representation is fixed and that it prescribes a well defined organisation of features, we expect the optimal solution $\hat\theta$ to be unique, whereby weights correspond to specific features of the data. 
}.
We resort to a classical variational argument which is at the basis of Expectation-Maximisation (EM) algorithms~\cite{neal1998view}. Consider the maximisation of the function
\begin{eqnarray}
F(Q,\theta) & = & \EQ{\log p(\bx,\bs|\theta)}+H[Q] \\
 & = & \log p(\bx|\theta)-D_{KL}\left( Q||p(\bs|\bx,\theta)\right)
\end{eqnarray}
over the parameters $\theta$ and over the distribution $Q(\bs|\bx)$. Here $\EQ{\ldots}$ stands for the expected value over the distribution $Q$. Inspection of the second line shows that the solution of this problem  is equivalent to the maximisation of the likelihood over the parameters $\theta$, because the optimum on $Q$ is at $Q(\bs|\bx)=p(\bs|\bx,\theta)$. 

If we expect that the solution $p(\bs|\bx,\theta)$ is sharply peaked around one value $\bs(\bx)$, we can search for a maximum on the set of distributions
\begin{equation}
\label{ }
Q(\bs|\bx)=\delta_{\bs,\bs(\bx)}.
\end{equation}
Within this class, the entropy of $Q$ is zero ($H[Q]=0$) and hence $F(Q,\theta)$ becomes the joint log-likelihood $\log p(\bx,\bs(\bx)|\theta)$. We shall then pursue the maximisation of the joint likelihood instead of the likelihood. This entails associating to each point $\bx^{(a)}$ of the sample an internal state $\bs^{(a)}=\bs(\bx^{(a)})$ in such a way as to maximise the joint log-likelihood
\begin{equation}
\label{jLL}
\tilde\mathcal{L}=\frac 1 N\sum_{a=1}^N \log p(\bx^{(a)},\bs^{(a)}|\theta)=\frac 1 N\sum_{a=1}^N \log p(\bx^{(a)}|\bs^{(a)},\theta)+\frac 1 N\sum_{a=1}^N \log p(\bs^{(a)}).
\end{equation}
As in EM algorithms, the optimisation can be done in two steps: in the E-step one internal state $\bs^{(a)}$ is associated to each data point $\bx^{(a)}$ so as to maximise $\tilde\mathcal{L}$ at fixed parameters $\theta$. In the M-step the parameters $\theta$ are adjusted in order to maximise $\tilde\mathcal{L}$ for fixed $\bs^{(a)}$. The M-step only involves the conditional distribution $p(\bx|\bs,\theta)$. Indeed the second term in the r.h.s. of Eq.~(\ref{jLL}) is independent of $\theta$. This term can be written as 
\begin{equation}
\label{jLL2}
\frac 1 N\sum_{a=1}^N \log p(\bs^{(a)})=-H[q]-D_{KL}\left(q(\bs)||p(\bs)\right)
\end{equation}
where $H[q]$ is the entropy of the {\em clamped state distribution}
\begin{equation}
\label{ }
q(\bs)=\frac 1 N \sum_{a=1}^N \delta_{\bs,\bs^{(a)}}
\end{equation}
i.e. the distribution which is induced by the data in the internal layer. The assignment of internal states $\bs^{(a)}$ in the the E-step then strikes a balance between describing optimally the data (first term in Eq.~\ref{jLL}) and maximising the likelihood of the internal states  (second term in Eq.~\ref{jLL}), which in turn entails a tradeoff between compression (first term in the r.h.s. of Eq.~\ref{jLL2}) and driving the clamped state distribution as close as possible to $p(\bs)$ (second term in the r.h.s. of Eq.~\ref{jLL2}). 

Let us now discuss how features are added sequentially. Given the solution (i.e. assignments and weights) of the maximisation in a machine with $n$ features, we start the dynamics assuming $w_{n+1,j}=0$. In the first E-step we assign the value $s_{n+1}^{(a)}=1$ to the least well described data-points (those with lower likelihood), leaving $s_{n+1}^{(a)}=0$ for all the others. Note that with $w_{n+1,j}=0$ the likelihood (first term in Eq.~\ref{jLL}) is unaffected. 
The number of data-points with $s_{n+1}^{(a)}=1$ is then determined by the last term in Eq.~(\ref{jLL}), in such a way that the fraction of sample points with $s_{n+1}^{(a)}=1$ matches the probability $p(\bs)=\xi^n/Z$ of states characterised by the $n+1^{\rm st}$ feature. In each M-step, we iterate gradient ascent steps until the maximal value of the gradients does not exceed $10^{-3}$ in absolute value. The algorithm stops when the assignment in the E-step does not change. We remark that the algorithm so defined is fully deterministic.

\section{Numerical experiments}
\label{sec:numerical}

In this Section we discuss the results of the algorithm that learns one feature of the HFM at a time discussed above. We shall briefly comment later on algorithms where features are learned all at once.
Our focus is on the properties of the representation as a function of the relevant parameters ($n$ and $g$) rather than on computational efficiency. 

\subsection{A synthetic dataset}

\begin{figure}
  \centering
\includegraphics[width=0.7\textwidth,angle=0]{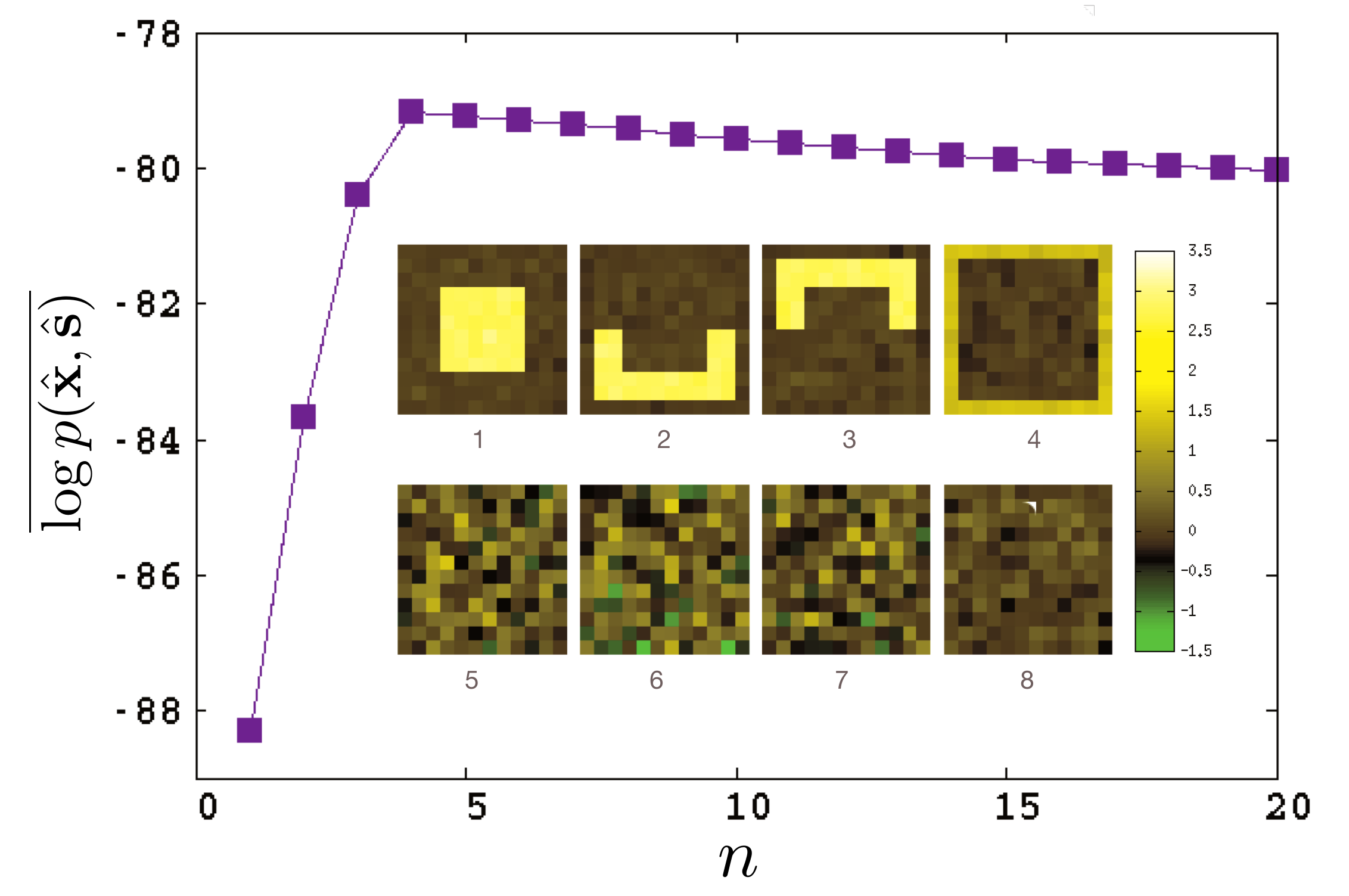}
  \caption{\label{fig:synth} Recovery of the structure of a HFM from a dataset of $N=6000$ images of $12\times 12$ pixels. The HFM used to generate the data contains four features, for each of these $w_{i,j}$ in $p(\bx|\bs)$ takes values $\log 4\simeq 1.39$ on a subset of the pixels and $w_{i,j}=0$ otherwise ($a_j=0$ for all $j$). This subset is the central square of $6\times 6$ pixels for the first feature, the lower (upper) half of the pixels at a distance 2 or 3 from the border for the second (third) feature, and the pixels on the outer perimeter for the fourth. The main plot shows the in-sample likelihood as a function of the number $n$ of features, for $g=0.70\simeq g_c$. The inset shows the features $\vec w_o$ recovered for $n=8$.} 
\end{figure}

We start by testing the ability of the algorithm to recover a structure of hierarchical features from a dataset generated by the HFM model itself. The data consists of $N=6000$ images of $12\times 12$ black or white pixels ($x_j=0$ or $1$) formed combining four features (see caption of Fig.~\ref{fig:synth}). Each image $\bx^{(a)}$ is generated drawing at random one internal state $\bs^{(a)}$ from $p(\bs)$ with $g=0.7\simeq g_c$, and then drawing $\bx^{(a)}$ from $p(\bx|\bs^{(a)})$. 

As shown, the joint likelihood sharply increases with $n$ when there are less than four features, but it decreases when $n>4$. The first four recovered features coincide with those used in the model. The features of order $i>4$ appear to be noisy and they become weaker and weaker moving down the hierarchy.

\subsection{rMNIST}

We trained the HFM on a reduced version of the MNIST dataset of hand written digits, that we call the rMNIST\footnote{Learning one feature at a time is not very efficient, which restricts the dimension of the dataset we could handle, given our computational resources.}. This dataset was used also in a previous publication~\cite{Odilon} and it consists of a training set of $N=60000$ images of $12\times 12$ black and white pixels ($x_j=0, 1$, $j=1,\ldots, 144$) obtained from the original $28\times 28$ MNIST grayscale images\footnote{We first transform the dataset into coarse grained pixels of $2\times 2$ original pixels, and binarise the result applying a threshold. Then we focus on the central image of $12\times 12$ pixels.}. A test set of $N=10^4$ images is obtained in the same manner from the MNIST dataset.

\begin{figure}
  \centering
\includegraphics[width=0.9\textwidth,angle=0]{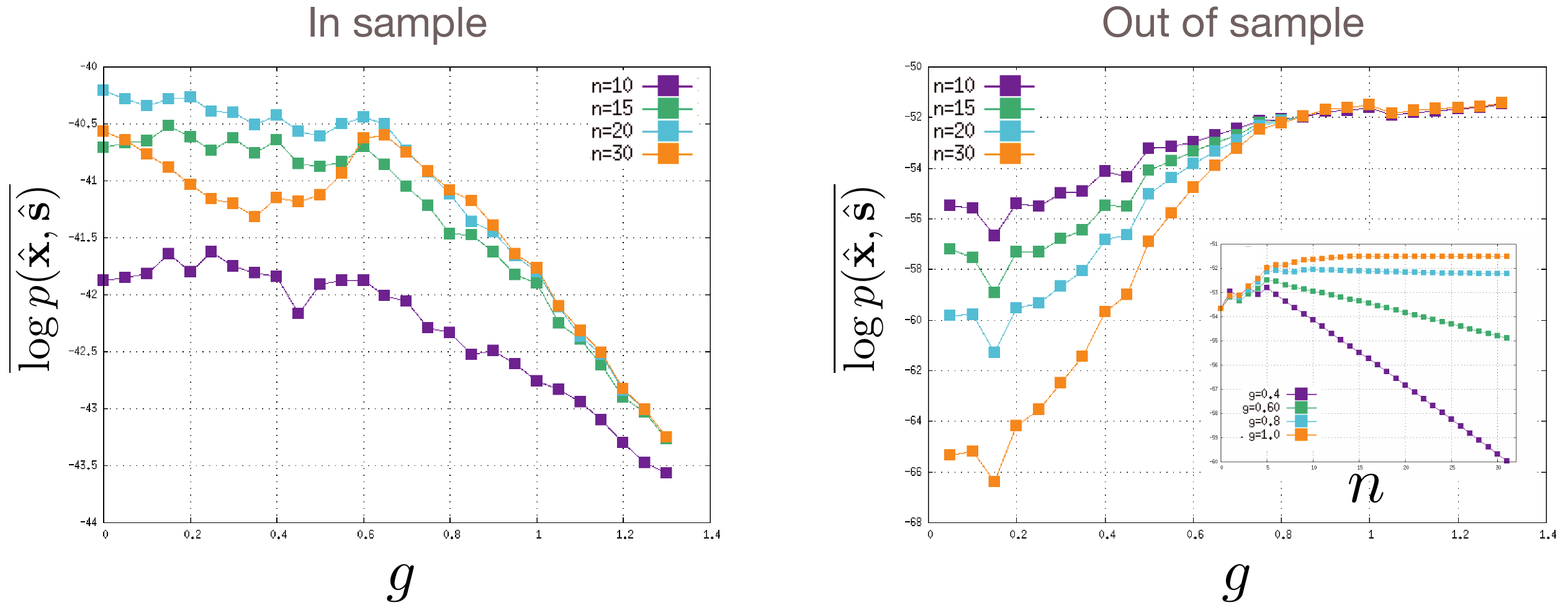}
  \caption{\label{fig:likMNIST} Joint log-likelihood in-sample (left) and out-of-sample (right) for the rMNIST dataset, for networks with different numbers ($n$) of hidden nodes, as a function of the parameter $g$. The inset in the left panel shows the out-of-sample joint log-likelihood as a function of $n$, for different values of $g$.} 
\end{figure}

Fig.~\ref{fig:likMNIST} shows the in-sample and out-of-sample likelihood\footnote{The in-sample likelihood is computed on the $N=60000$ examples of the training set, whereas the out-of-sample likelihood is computed on $10^4$ test set digits.} for the 
rMNIST dataset  as a function of $g$ for different values of $n$. These have opposite behaviours: while the in-sample likelihood generally decreases with $g$, the out-of-sample likelihood increases. In both cases, the behaviour changes qualitatively when $g\approx g_c\simeq 0.7$. For $g>g_c$, the out-of-sample likelihood levels off to a nearly constant value, for all $n$. In the same region, the in-sample likelihood decreases sharply with $g$, whereas it hovers around a constant value for $g<g_c$. 

The inset shows that for $g<g_c$ the statistical accuracy degrades upon adding more features, suggesting that the model is overfitting the data. For $g>g_c$ instead the out-of-sample log-likelihood  levels off to a constant value as $n$ increases. This is related to the fact that the effective number of features $\mathcal{H}$ is finite for $g>g_c$ and it increases linearly with $n$ for $g\le g_c$.

\begin{figure}
  \centering
\includegraphics[width=0.9\textwidth,angle=0]{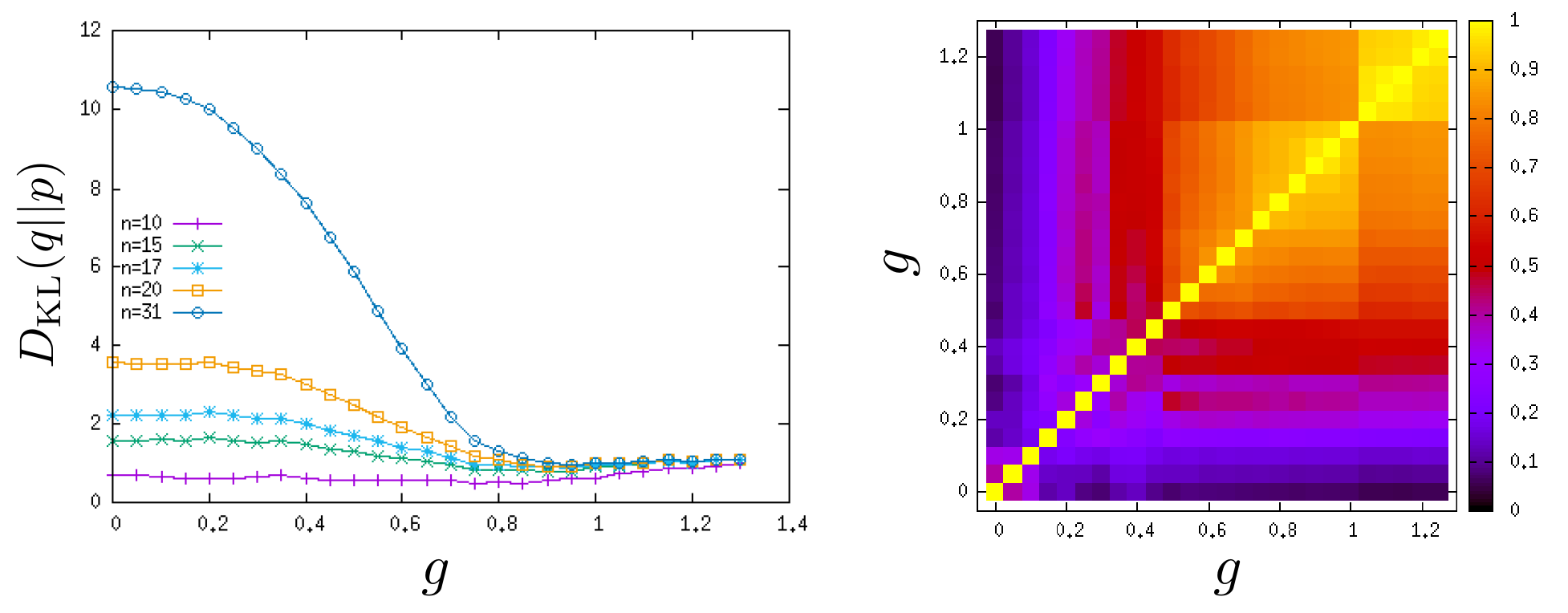}
  \caption{\label{fig:dkl} Left: Kullback-Leibler divergence between the final (empirical) distribution $q(\bs)$ after training, and the initial one $p(\bs)$, for the rMNIST dataset. Right: overlap between the internal representations of the rMNIST dataset for different values of $g$. The overlap between two configurations $\bs^{(a)}_\alpha$ and $\bs^{(a)}_\beta$ is computed as $o_{\alpha,\beta}=\frac{1}{nN}\sum_{i,a}s_{i,\alpha}^{(a)}s_{i,\beta}^{(a)}$, where $s_{i,\alpha}^{(a)}$ is the value of the $i^{\rm th}$ variable of the internal representation of the $a^{\rm th}$ data point $\bs^{(a)}_\alpha$.}
\end{figure}

Fig.~\ref{fig:dkl} provides a complementary perspective on the trained HFM. First (left panel of Fig.~\ref{fig:dkl}) it shows that the occupation of internal states $\bs$ is very far from the equilibrium distribution $p(\bs)$ for $g<g_c$ whereas $q(\bs)$ stays close to $p(\bs)$ for $g>g_c$. 

We expect that the model exhibits a poor generative performance when $g<g_c$, because the distribution $q(\bs)$ induced by the data is far from $p(\bs)$, and it is expected to improve for $g>g_c$.
The right panel of Fig.~\ref{fig:dkl} shows that the occupation of internal states changes smoothly for $g>g_c$ but it changes abruptly for $g<g_c$. 

Taken together, these observations suggest that the optimal results are obtained when the parameter $g$ is close to its critical value.

\begin{figure}
  \centering
\includegraphics[width=0.9\textwidth,angle=0]{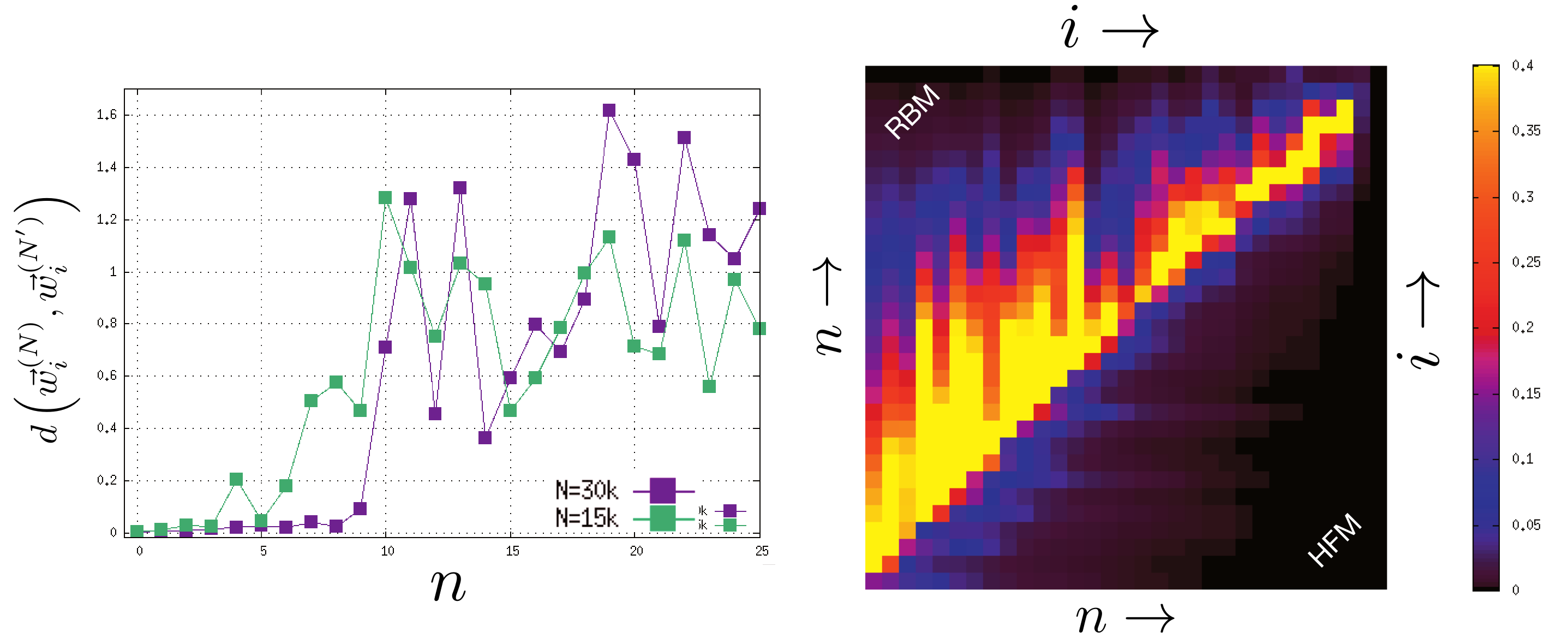}
  \caption{\label{fig:distw} Left: distance $d(\vec w_i,\vec w_i')$ between the $i^{\rm th}$ feature $\vec w_i$ learned with $N$ data points and the one ($\vec w_i'$) learned with the whole rMNIST dataset ($6\cdot 10^4$ data points). The green points correspond to $N=1.5\cdot 10^4$, the violet ones to $N=3\cdot 10^4$. Right: Distance $d(\vec w_i^{(n)},\vec w_i)$  between the $i^{\rm th}$ feature $\vec w_i^{(n)}$ in a machine with $n$ features and the $i^{\rm th}$ feature $\vec w_i$ in a machine with $31$ features. The lower triangular data refer to the HFM ($i$ increases from bottom to top and $n$ from left to right). The upper triangular data refer to a RBM ($i$ increases from  left to right and $n$ from bottom to top).}
\end{figure}

The features learned in the HFM converge quickly to an asymptotic value when more features or more data-points are added. This is confirmed by Fig.~\ref{fig:distw} (left), which plots the distance
\begin{equation}
\label{ }
d(\vec{w}_i,\vec{w}_i')=1-\frac{\vec w_i\cdot\vec w_i'}{|\vec w_i||\vec w_i'|}
\end{equation}
between the $i^{\rm th}$ feature $\vec w_i$ learned with $N$ data-points and the same feature $\vec w_i'$ learned with the whole dataset ($N=60000$). This suggests that low order features are learned first (i.e. with few data points) whereas higher order ones need more data to converge. 

A similar analysis is reported in the right panel of Fig.~\ref{fig:distw}, when $N$ is fixed but the number $n$ of features changes. For the HFM the distance of $\vec w_i$ when $n$ features are learned to its value when the final number of features are learned is shown in the part below the diagonal of the plot in the right panel of Fig.~\ref{fig:distw}. For comparison we show the same plot for a RBM in the upper diagonal part (with axes inverted) trained on the same data\footnote{We train RBMs for 
for $300$ epochs, with persistent contrastive divergence and mini-batches of $10$ data-points, as in Ref.~\cite{Odilon}.}. For the RBM the introduction of a new feature perturbs the already learned ones, for the HFM low order features remain stable even when further high order features are learned.

\begin{figure}
  \centering
\includegraphics[width=0.45\textwidth,angle=0]{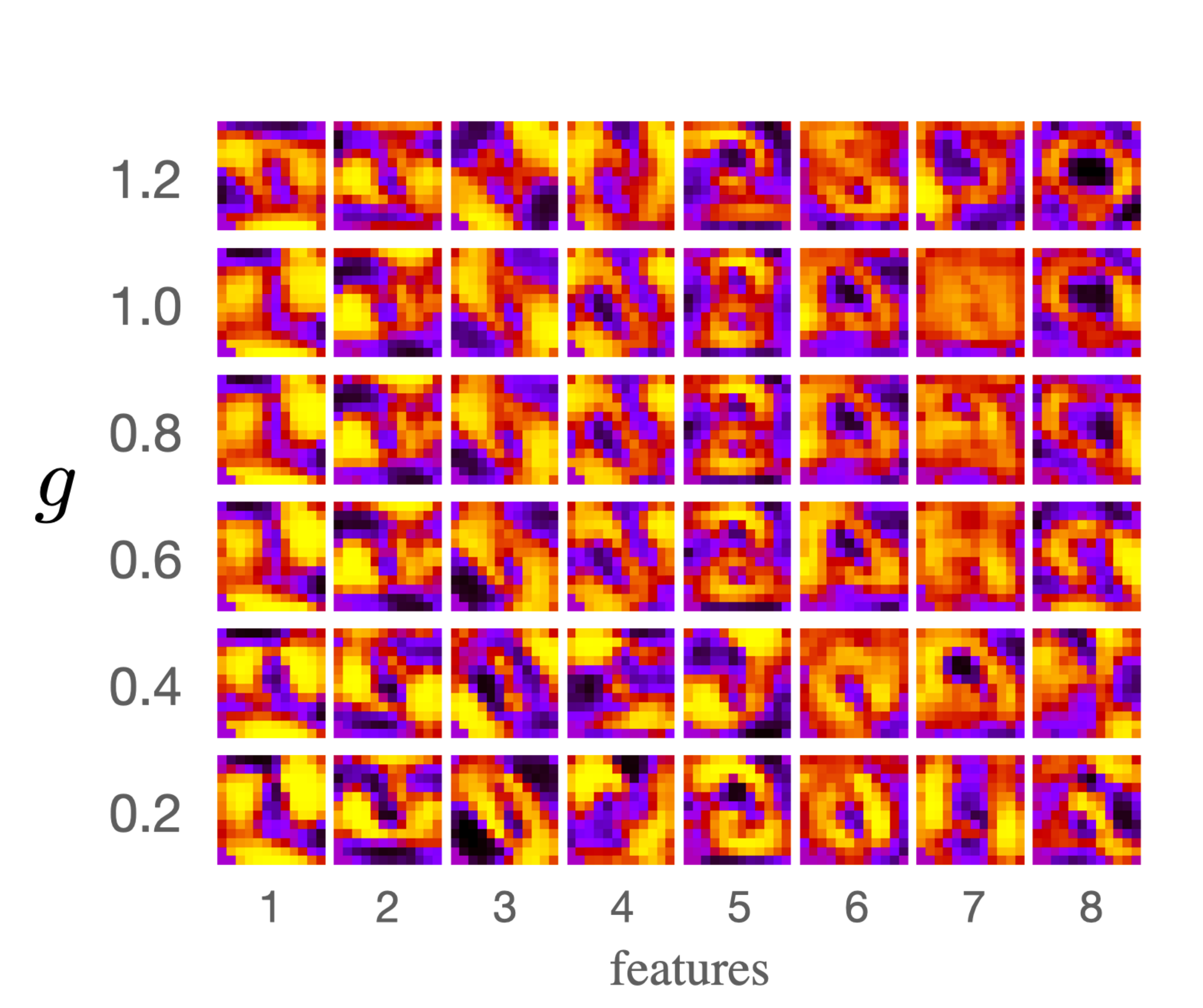}
\includegraphics[width=0.45\textwidth,angle=0]{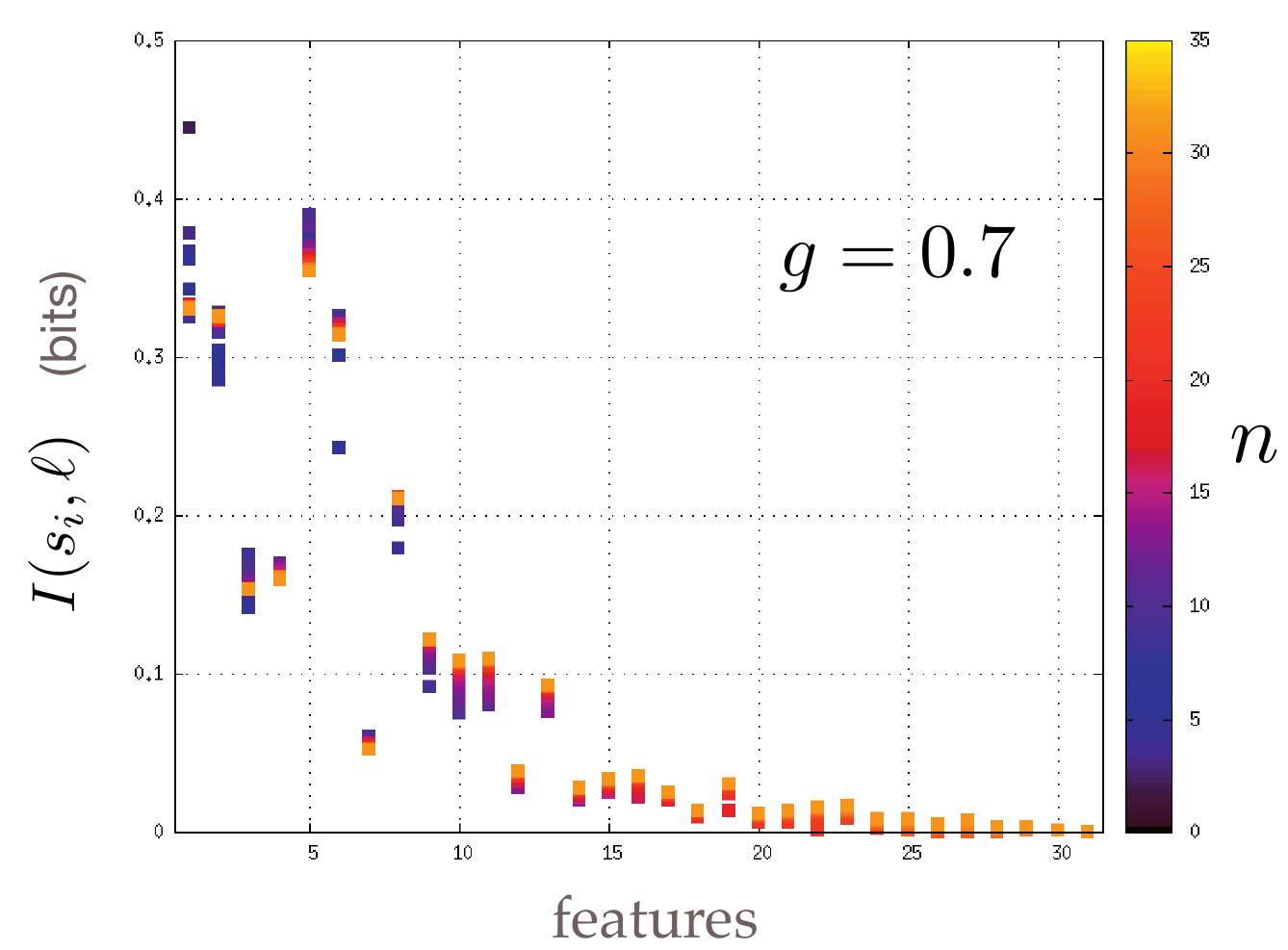}
  \caption{\label{fig:features} Left: the first $8$ features of a HFM with $n=31$ features, for different values of $g$. Each pixel reports the value of $(1+e^{-w_{i,j}})^{-1}$ so that its value is between zero (Black) and one (yellow). Right: Mutual information between labels $\ell$ and the $i^{\rm th}$ feature $s_i$, for $g=0.7$. Different values of $n$ correspond to different colours (see scale on the right).}
\end{figure}
The robustness of the internal representation also manifests in the fact that features change very little as the parameter $g$ varies. This is shown in Fig.~\ref{fig:features}. The hierarchical organisation of features is also evident in the fact that features become weaker and weaker as their order increases. 
A further evidence of this, is that relevant information on the data, such as the value of the digits in the rMNIST case, is mostly contained in low order features, as shown in Fig.~\ref{fig:features} (right).

\subsection{The Big Five personality test}

The Big Five Personality Test (Big5)\footnote{Available at {\tt https://www.kaggle.com/datasets/tunguz/big-five-personality-test}.} is a questionnaire with $m=50$ questions that are believed~\cite{Big5} to probe the personality of individuals along five traits: extraversion, agreeableness, openness, conscientiousness, and neuroticism. Respondents give a score from $1$ to $5$ to each question, which we rescale to the interval $[0,1]$. The dataset is composed of $N=1015342$ compiled questionnaires.

\begin{figure}
  \centering
\includegraphics[width=0.9\textwidth,angle=0]{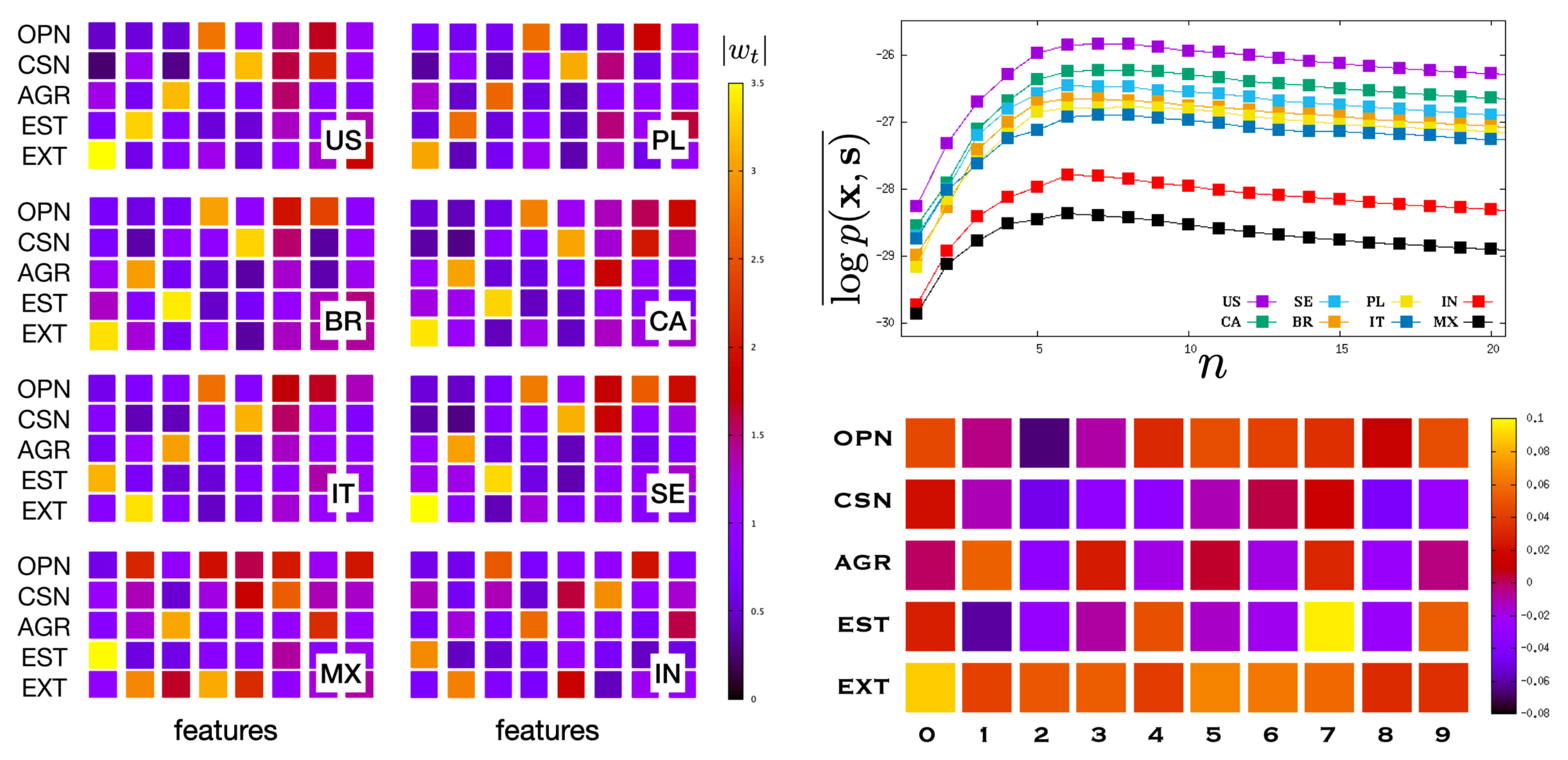}
  \caption{\label{fig:Big5} Left: Average size of the weights $w_{i,j}$ corresponding to the five blocks of ten questions relative to the five traits: openness (OPN), conscientiousness (CSN), agreeableness (AGR), neuroticism (EST) and extraversion (EXT). The 8 panels show that the top five features coincide with the first five traits for all countries except India (IN) and Mexico (MX). The figure displays on a colour scale the value the average weight strength $|w_{i,t}|=\sqrt{\frac{1}{10}\sum_{j\in t}w_{i,j}^2}$ that feature $i$ attributes to questions related to one of the traits $t$. The number of data points for the different countries is $N=472320, 5481, 4570, 9389, 4008, 53663, 9122$ and $13899$ for United States (US), Brazil (BR), Italy (IT), Mexico (MX), Poland (PL),  Canada (CA), Sweden (SW) and India (IN). Right top: In-sample joint log-likelihood for the same datasets as a function of the number $n$ of features, for $g=g_c$. Right bottom: Correlation between digits and personality traits. We compute the expected value of $x_j$ for the US Big Five dataset, conditioned to internal states that correspond to a specific digit in the MNIST database. We average the excess $\langle x_j|\ell\rangle-\langle x_j\rangle$ output for each digit over the $10$ questions relative to each trait. Each square in the bottom right panel reports the value of the excess output 
$\delta x_{t,\ell}=\frac{1}{10}\sum_{j\in t}  \langle x_j|\ell\rangle-\langle x_j\rangle$ for each trait and for each digit $\ell=0,1,\ldots,9$.}
\end{figure}

We run the algorithm described in the previous Sections on the Big Five dataset separately for respondents of different countries. In all cases we find that the log-likelihood reaches a maximum at six or seven features and then it decreases if more features are added (see Fig.~\ref{fig:Big5} right top). This suggests that the optimal number of features is close to five. Closer inspection of the features shows that, for all countries except two (India and Mexico), each of the first five features significantly concentrates on the sets of ten questions that correspond to one of the five traits. The two countries where the correspondence between features and traits is not sharp (India and Mexico) also exhibit lower values of the log-likelihood. The ranking of the traits agrees across some countries (US and PL or BR, CA and SW) and it differs in others. A closer analysis of the significance of these results goes beyond the scope of this paper. The main purpose of this exercise is to show that the HFM uncovers a clear structure of features, confirming the relevance of the five traits but also providing further insight into the variation of the organisation of the trait space across countries.

An architecture with a fixed internal representation makes it possible to establish a relation between data learned in different contexts. This type of associations is reminiscent of the phenomenon of synesthesia~\cite{watson2014synesthesia} in psychology, where the stimulation of one sensory pathway (e.g. grapheme) evokes a totally unrelated perception (e.g. colour). 

Fig.~\ref{fig:Big5} (bottom right) illustrates this for the case of digits and personality traits. It plots the excess output $\langle x_j|\ell\rangle-\langle x_j\rangle$ averaged over questions that probe one of the traits, where the conditional average is estimated as
\begin{equation}
\label{Correla}
\langle x_j|\ell\rangle
=\frac{1}{N_\ell}\sum_{a:\ell^{(a)}=\ell} p(x_j=1|\bs^{(a)}_{\tiny {\rm MNIST}},\hat \theta_{\tiny {\rm Big~Five}})
\end{equation}
and $\langle x_j\rangle=\frac{1}{10}\sum_{\ell=0}^9 \langle x_j|\ell\rangle$.
In Eq.~(\ref{Correla}) the sum runs on the $N_\ell$ rMNIST samples which correspond to digit $\ell$, $\bs^{(a)}_{\tiny {\rm MNIST}}$ is the hidden states for sample $a$ in the rMNIST dataset, and $\hat \theta_{\tiny {\rm Big~Five}}$ are the parameters of the network trained on the Big Five dataset. 

The most significant deviations from the baseline are observed for neuroticism, which is enhanced for $\ell=7$ and suppressed for $\ell=1$, for openness which is suppressed for $\ell=2$, and for extraversion which is enhanced for $\ell=0$. We refrain from commenting these results further. The aim of this exercise is to illustrate how architectures where the internal representation is fixed, makes it possible to establish correlations between data learned in different contexts, which have no obvious {\em a priori} relation. 

\subsection{Learning features all at once}
\label{sec:universalps}

\begin{figure}
  \centering
\includegraphics[width=0.7\textwidth,angle=0]{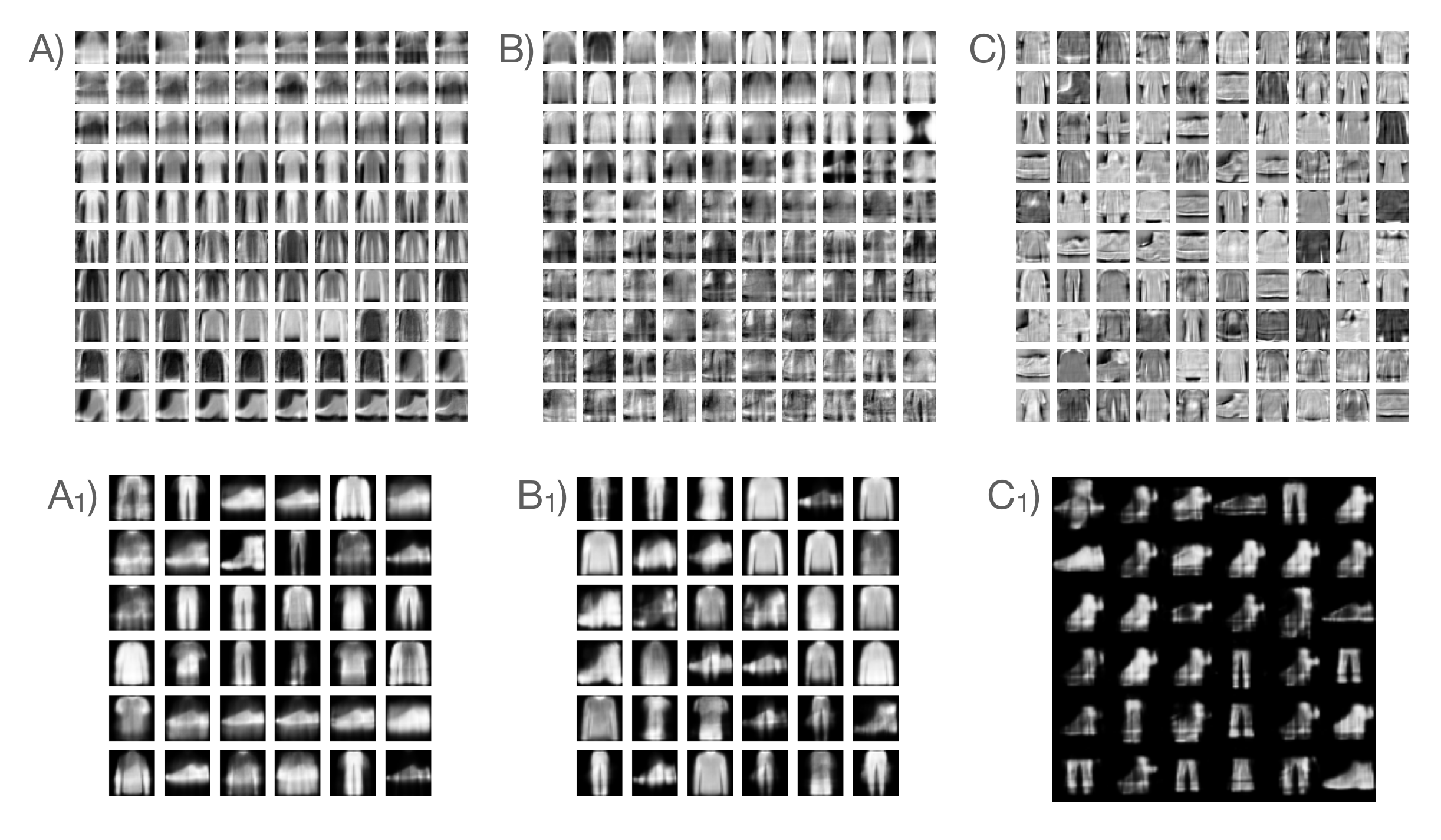}
  \caption{\label{fig:all_feat} Left: learned weights $w_{i,j}$ (A) and generated images (A${}_1$) for a network based on the HFM at $g_c$ with $n=100$ features, learned all at once from random initial conditions. The first feature is the one on the top left corner, the second is the one on its right and so on. Centre: weights $w_{i,j}$ (B) and generated images (B${}_1$) for the same model learning starts from the learned features of the same model with $n'=50$ (top $50$ features). The bottom $50$ features are zero when training starts. Right: weights $w_{i,j}$ (C) learned in a RBM trained for $100$ epochs with Contrastive Divergence with $k=50$, and generated images (C${}_1$) after $300$ Monte Carlo steps starting from a random initial condition. In all experiments, networks were trained on the EMNIST dataset.}
\end{figure}

Up until now, we learned progressively features one at a time. In order to understand how the results depend on this choice, we  explored algorithms where features are learned all at once. The algorithm we develop maximises again the joint likelihood. It starts by drawing the internal state $\bs^{(a)}$ at random from $p(\bs)$ and it proceeds alternating an M-step where it  maximise the likelihood over the parameters by gradient ascent and an E-step where it modifies the assignment $\bs^{(a)}$ to maximise the likelihood. We provide more information on the algorithm in Appendix~\ref{app:algo}.

Figure~\ref{fig:all_feat} shows the results of numerical experiments of the algorithm with $n=100$ features (left panel) on the Fashion MNIST database\footnote{The Fashion MNIST datasets is a collection of $N=6\cdot 10^4$ grayscale images of $28\times 28$ pixels ($m=784$), that consists of Zalando's article images~\cite{zalando} and it is available at {\tt https://github.com/zalandoresearch/fashion-mnist}.}. This is compared to the results obtained when $n=50$ features are first learned (keeping $w_{i,j}=0$ for all $i>50$) and then the remaining features are learned, starting from the features learned for $n=50$ (central panel). Both experiments are done at $g=g_c$.
The features in the two cases differ substantially: In the second case, the first features $50$ that are learned appear sharper than the $50$ features that are learned later. When all $100$ features are learned together, we observe several features that are close duplicates of other ones. This suggests a natural way in which the model could be expanded in the over-parametrised regime, by duplicating already existing features. This is consistent with the peculiar structure of the HFM which, at $g=g_c$, assigns the same relative probability to states with $k$ features in the original model and to states with $2k$ features in the model with duplicated features. As discussed in  Appendix~\ref{app:HFM} this property derives from $g_c$ being a fixed point in a renormalisation group transformation. 
The generated images in A${}_1$ and B${}_1$ show that the excess redundancy allows the network to generate better data. 

The leftmost panel reports the same experiment run on an RBM with $n=100$ hidden nodes trained with contrastive divergence ($k=50$). As observed in Ref.~\cite{decelle2021equilibrium}, the generation of images starting from random initial conditions is problematic. Indeed even after $300$ Montecarlo steps, the generated images (C${}_1$) are not very good. Notice that while the generated images in A${}_1$ and B${}_1$ are equilibrium samples from the distribution $p(\bx|\hat\theta)$, the same cannot be said for panel C${}_1$, because sampling from the equilibrium distribution of RBMs is non-trivial, as discussed in Ref.~\cite{decelle2021equilibrium}. The comparison with 
panels A and B shows that the features extracted by the RBM (panel C) lack the particular organisation among features of the HFM. 

\section{Discussion} 

Motivated by an interpretation of the term ``understanding'' as mapping data to an already existing representation, this paper proposes a simple model of a probabilistic single layer networks with a fixed internal representation, which 
is derived from the principle of maximal relevance~\cite{marsili2022quantifying}. The previous Sections illustrate the properties of this learning modality with applications to few datasets. In this concluding Section, we shall first contrast this learning modality with the most traditional one, and conclude with some remarks on future work. 

\subsection{Two learning modalities}

There are several qualitative differences between an approach to learning with a fixed internal representation and one where $p(\bs)$ is learned from data. We discuss these differences taking RBMs as a paradigm for the latter. Appendix~\ref{app:RBM} briefly recalls the definition and the main properties of RBMs.
\begin{description}
  \item[Hierarchy]  The HFM imposes a hierarchy between features {\em a priori} whereas the different hidden variables $s_i$ in an RBM are all a priori equivalent, with no hierarchical organisation of relevance among them (see~\cite{cote2016infinite} for a version of RBMs with hierarchical features). 
As a consequence, relevant information (e.g. on the labels of a dataset) is scattered on all hidden nodes of a RBM, as observed by Fernandez~de~Cossio~Diaz {\em et al.}\footnote{Fernandez~de~Cossio~Diaz {\em et al.}~\cite{fernandez2022disentangling} also show how the training process can be modified in order to concentrate relevant information on few informative nodes.}~\cite{fernandez2022disentangling}, whereas it concentrates on the top features of the HFM (see Fig.~\ref{fig:features} right). Although a hierarchy of features is not imposed {\em a priori} in RBMs, Decelle and Furtlehner~\cite{decelle2021restricted} argue that it emerges during training in the singular value decomposition of weights.
  
  \item[Uniqueness and determinism] The learned internal representation $p(\bs|\hat\theta)$ in a RBM is not unique, but it  belongs to a manifold of possible distributions. The learned representation to which training converges depends on the (random) choice of initial conditions and on the specific realisation of the stochastic dynamics\footnote{Training in neural networks generally relies on stochastic gradient ascent dynamics, whereby the gradients of the likelihood are estimated on random subsamples of the dataset (mini-batches)~\cite{hinton2012practical}. It is worth mentioning that Moschella {\em et al.}~\cite{moschella2022relative} suggest that randomness does not affect the internal representation of neural networks, when this is expressed in terms of suitable relative coordinates.}. By construction, when the internal representation $p(\bs)$ is fixed, it is also unique. This decreases drastically the complexity of the likelihood landscape that the training dynamics has to go through, to the point that, as we have shown, training can be deterministic without the need to introduce stochasticity. 
  
\item[Complexity] When $p(\bs)$ is fixed, it is possible to choose the internal representation according to Occam's razor, as a very simple model (e.g. the HFM)\footnote{Simplicity can be quantified in terms of Minimum Description Length. In this framework, the leading term in the complexity is proportional to $\frac K 2 \log N$, where $K$ is the number of parameters~\cite{myung}. $K=1$ for the HFM whereas $K=nm+n+m$ for the distribution $p(\bs|\theta)$ of the RBM.}. This is consistent with common wisdom in classical statistical inference, whereby too complex models generally over-fit the data, i.e. have poor generalisation performance~\cite{myung}. 

Neural networks such as RBMs do not obey this logic. Their performance does not degrade as the number of parameters (i.e. of hidden variables) grows large. Indeed there is mounting evidence that learning in over-parametrised neural networks follows different rules~\cite{zhang2021understanding}, which we only partially understand (see e.g.~\cite{mei2019generalization}). 

It has been recently pointed out~\cite{decelle2021equilibrium} that the complexity of the energy landscape of RBMs makes it very costly to learn the true equilibrium distribution. On the contrary, when $p(\bs)$ is chosen {\em a priori} learning converges smoothly to the equilibrium distribution and sampling from it is very easy.

\item[Generic vs specific features; plasticity] By construction, when $p(\bs)$  is fixed {\em a priori}, the internal representation is independent of the data. All the dependence on the data is necessarily stored into the feature vectors $\vec w_i$. Therefore, features carry data-specific information whereas $p(\bs)$ does not. When $p(\bs)$  is fixed, learning amounts to feature extraction. 

The situation is very different when $p(\bs|\hat \theta)$ is learned during training, as in RBMs. Then training converges to one of many possible solutions (see above) which implies that features are necessarily generic, i.e. they depend weakly on the data (see e.g.~\cite{decelle2021restricted,ingrosso2022data}). This has the advantage that the weights learned from one dataset can reasonably well reproduce inputs drawn from a different dataset. This is illustrated in  Fig.~\ref{fig:emnist} (see caption) that shows 
outputs that correspond to states $\bs_E$ clamped on EMNIST images\footnote{The EMNIST datasets are collections of grayscale images of $28\times 28$ pixels ($m=784$) of handwritten characters (lowercase letters in our case). We normalize each pixel into the range of $0$ to $1$. The dataset is available at {\tt https://www.nist.gov/itl/products-and-services/emnist-dataset}.}, of an RBM trained on MNIST. The states $\bs_E$ are very atypical, with a probability $p(\bs_E)$ which is typically a factor $\sim 10^{-52}$ smaller than that of the clamped hidden states corresponding to the MNIST dataset with which the RBM was trained. This property is key for transfer learning\footnote{Transfer learning is a set of applications in supervised learning where networks trained on a given dataset are used for supervised learning tasks on different datasets, by retraining only the output layer.}~\cite{weiss2016survey}. It also suggests low degrees of ``plasticity", i.e. that weights should change by a small amount when the RBM is further trained on a different dataset. Data specific information is stored in $p(\bs|\hat \theta)$, which is a compressed representation of the data. This suggests that learning in RBMs is similar to data compression, whereby a data source $p(\bx)$ is transformed into a compressed representation $p(\bs|\hat\theta)$, using a dictionary of codewords -- the weights\footnote{Note indeed the similarity: optimal data compression~\cite{CoverThomas} also does not have a unique solution and codewords are generic.}. 

\begin{figure}
  \centering
\includegraphics[width=0.7\textwidth,angle=0]{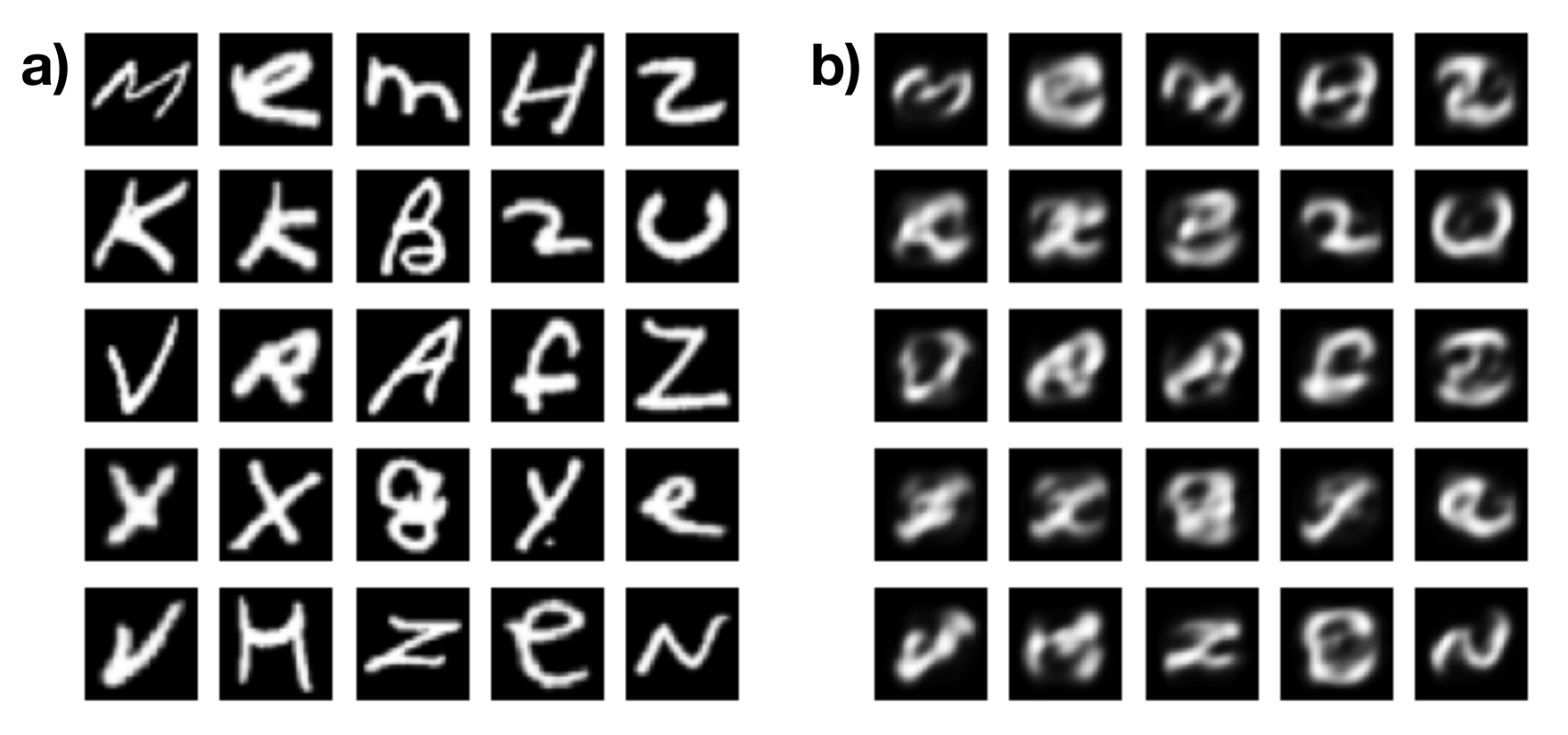}
  \caption{\label{fig:emnist} a) A sample of characters from the EMNIST dataset. b) output of a RBM trained on the MNIST dataset when clamped on the EMNIST characters in a). These are obtained finding the most likely internal state $\bs_E={\rm arg}\max_{\bs} p(\bx_E,\bs)$ of an RBM where the visible layer is fixed to an EMNIST character $\bx_E$. In b) we show the plot of $\E{\bx|\bs_E}$ generated by the RBM. The RBM has $n=100$ hidden variables and is trained on MNIST for $100$ epochs with mini-batches of $50$ data points and contrastive divergence with $k=100$. The clamped states $\bs_E$ which correspond to EMNIST characters $\bx_E$ are very atypical: Their probability $p(\bs_E)$ is typically a factor $\sim 10^{-52}$ smaller than that of the clamped hidden states corresponding to the MNIST dataset with which the RBM was trained with.}
\end{figure}

  \item[Continuity and compressibility] When $p(\bs)$ is fixed, the learned features change smoothly when new features or new data are added, or when $g$ is varied (see Fig.~\ref{fig:distw}). The HFM provides a precise organisation to the feature space in which new features can be introduced without affecting already learned ones.
This endows learning with a degree of flexibility that makes it possible, for example, to tune the resolution $H[\bs]$ of the internal representation continuously by changing $g$. 
  
By contrast, the resolution $H[\bs]$ of the internal representation in a RBM is determined by the number of hidden nodes and by the structure of the data in a complex manner, and it cannot be easily tuned. The weights and the internal representation of a trained RBM vary in a discontinuous manner when new hidden nodes are added (see Fig.~\ref{fig:distw} right). 
Because of the stochastic nature of the training process, upon adding further data points to the dataset (i.e. $N\to N'>N$) the learned representation may also change. 

\item[Beyond generalisation] Architectures where $p(\bs)$ is fixed allow for tasks that go beyond reproducing data or generalising. For example, once the conditional distributions $p(\bx|\bs,\hat\theta)$ and $p(\bz|\bs,\hat\theta')$ are learned from two different datasets $\hat\bx$ and $\hat\bz$, it is possible to compute a joint distribution
\begin{equation}
\label{ }
p(\bx,\bz)=\sum_{\bs}p(\bx|\bs,\hat\theta)p(\bz|\bs,\hat\theta') p(\bs)
\end{equation}
that may shed light on possible relations between $\bx$ and $\bz$, as shown in Fig.~\ref{fig:Big5} (bottom right). Drawing such hypothetical relations is a key aspect of intelligent behaviour. We speculate that architectures with a fixed $p(\bs)$ can generate combination of the learned features which are not observed in the data, an ability akin to imagination~\cite{mahadevan2018imagination}.
\end{description} 

\subsection{Outlook}

Taken together, these observations suggest that unsupervised learning in architectures where $p(\bs)$ is fixed {\em a priori} is qualitatively different from learning when the internal representation is learned, as in RBMs. Learning in RBM type architectures is similar to data compression, whereby a data source $p(\bx)$ is transformed into a compressed representation $p(\bs|\hat\theta)$ by generic weights. When the internal representation is fixed, the learned weights are very specific of the data with which the model is trained. Therefore learning is akin to feature extraction or to ``understanding''.

Architectures where the internal representation is shaped during training, such as RBMs, are reminiscent of the empiricist approach to learning, that starts from a blank slate initial state and all knowledge comes from experience. By contrast, architectures with a fixed internal representation are reminiscent of nativism approaches, which assume an innate ``universal structure'', such as the one hypothesised by Chomsky for language~\cite{chomsky1976reflections}. It is tempting to conclude that both architectures are used in more complex learning machines such as the brain.
Early stages of information processing in sensory systems, such as the retina or the primary visual cortex in vision, are  adapted to process a large variation of structured datasets. These areas encode representations in terms of generic features, such as localised filters~\cite{olshausen1996emergence}, and they exhibit suppressed levels of experience dependent plasticity after development~\cite{berardi2000critical}. This is reminiscent of the RBM type learning modality, as discussed above. 

At the same time, architectures where the internal representation is fixed store data specific information in the weights as in models of associative memories~\cite{hopfield1982neural}. We also argued that these networks enable functions that go beyond the mere generalisation -- such as generating objects which have never been seen or drawing analogies between data learned in different domains -- which are associated to higher cortical areas. 

This points to the intriguing hypothesis that the nature of internal representations evolves with depth, with early steps of information processing that transform data in compressed representations using generic features, and deeper layers that extract features which are stored in the weights, within a data independent internal representation. In this perspective, 
understanding whether a ``universal'' (data-independent) internal representations emerges spontaneously in deep layers of learning architectures is an interesting avenue for further investigation.

A related avenue of research concerns relating these findings to thermodynamic costs of learning, drawing from recent advances in stochastic thermodynamics~\cite{parrondo2015thermodynamics,still2020thermodynamic}. If, as argued above (see footnote~\ref{foot:thermo}), thermodynamic costs are related to the compression of the state space of the internal representation, it makes sense to learn over-parametrised generic representations such as those of RBMs once and for all for a broad range of datasets, as done in the early areas of processing of visual stimuli~\cite{berardi2000critical}. On the contrary, when new information needs to be stored, the same thermodynamic argument suggests that learning should proceed by expanding the state space -- as in the algorithms described in this paper -- relying on simple models, such as the HFM. We defer further discussion along this line to future publications.


\bibliographystyle{unsrt} 
\bibliography{occam_learning.bib}

\section{Acknowledgments}

MM thanks Iacopo Mastromatteo who brought to our attention the HFM as a model of maximal relevance and Susanne Still for enlightening discussions on thermodynamics of learning (see footnote~\ref{foot:thermo}). We thank Roberto Mulet for interesting discussions in the initial stages of this work, Alessandro Ingrosso and Davide Zoccolan for insightful comments.
Rongrong Xie acknowledges a fellowship from the China Scholarship Council (CSC) under Grant CSC No. 202006770018.

\appendix

\section{The statistical mechanics of the HFM}
\label{app:HFM}

The derivation of the properties of the HFM becomes easy if all sums (and expected values) over $\bs$ are computed conditioning on the value of the energy. Let $\mathcal{S}_k=\{\bs:~m_{\bs}=k\}$ be the set of states with the most detailed feature being the $k^{\rm th}$ one. Then, for example, the partition function in Eq.~(\ref{eqZ}) is computed as
\[
Z=\sum_{k=0}^n\sum_{\bs\in\mathcal{S}_k}e^{-g\max\{k-1,0\}}=1+\sum_{k=1}^n\left(2e^{-g}\right)^{k-1}
\]
The other formulas in the main text are easily derived with the same method. Likewise, expected values over $\bs$ can be computed observing that
\begin{equation}
\label{ }
\E{s_\ell| \bs\in\mathcal{S}_k}=\left\{\begin{array}{rr}0 & ~\ell>k \\+1 & ~\ell=k \\1/2 & ~\ell<k \end{array}\right.\,.
\end{equation}
Therefore
\begin{equation}
\label{ }
\E{s_\ell}=\sum_{k=0}^n P\{\bs\in \mathcal{S}_k\} \E{s_\ell| \bs\in\mathcal{S}_k}=\frac 1 2 \left[1+\frac{(\xi^{\ell-1}-1)(\xi-2)}{\xi^n+\xi-2}\right]
\end{equation}
(recall that $\xi=2e^{-g}\le 2$). At $\xi=1$ $\E{s_\ell}=\frac{n-i+2}{2(n+1)}$ varies linearly between $1/2$ and $1/(n+1)$. Also note that $\E{s_1}=\frac 1 2 $ for all $\xi\in [0,2]$ and that $\E{s_\ell}=\frac 1 2$ for $\xi=2$ (i.e. the non-interacting case $g=0$) for all values of $\ell$, as it should. 

The same logic can be applied to compute expected values of products of any number $m$ of variables, leading to the result 
\begin{equation}
\label{ }
\E{s_{\ell_1}\cdots s_{\ell_m}} =2^{1-m}\E{s_{\ell_m}}, \qquad (\ell_1<\ell_2<\ldots<\ell_m)\,,
\end{equation}
because the expected value only depends on the variable with the highest index. So, the connected correlation function is
\begin{equation}
\label{ }
C_{i,j}=\E{s_j} \left(\frac{1}{2} -\E{s_i}\right),\qquad i< j 
\end{equation}
whereas $C_{i,i}=\E{s_i} \left( 1  -\E{s_i}\right)$, as for any binary variable. Notice that $C_{1,j}=0$ for all $j>1$, because $s_1$ is independent of all $s_j$ with $j>1$.

This structure of correlations arises from a model that features interactions of arbitrary order. We can see this by writing  
\[
p(\bs)=e^{\sum_{\mu}g^\mu\phi^\mu(\bs)},\qquad \phi^\mu(s)=\prod_{i\in\mu}(2s_i-1)
\]
as an exponential model of spin variables $\sigma_i=2s_1-1\in\{\pm 1\}$, which depend on product operators $\phi^\mu$, as in Ref.~\cite{beretta2018stochastic}. Here $\mu$ is an integer between $0$ and $2^n-1$ and the product on $i\in\mu$ runs over all integers that correspond to a $1$ in the binary representation of $\mu$. Hence $\phi^0(\bs)=1$ is a constant and $g^0$ is a free energy term that depends on all $g^\mu$ for $\mu>0$ and ensures normalisation. 

As shown in Ref.~\cite{beretta2018stochastic}, the coefficients $g^\mu$ are given by the inversion formula
\begin{eqnarray*}
g^\mu & = & \frac{1}{2^n}\sum_{\bs}\phi^\mu(\bs)\log p(\bs) = -\frac{1}{2^n}\sum_{\bs}\phi^\mu(\bs)g\max(m_\bs-1,0)\\
 & = & -\frac{g}{2^n}\sum_{k=0}^n \max(k-1,0) \sum_{\bs\in \mathcal{S}_k} \prod_{i\in\mu} (2s_i-1)=
 -\frac{g}{2^n}\sum_{k=2}^n (k-1) \sum_{\bs\in \mathcal{S}_k} \prod_{i\in\mu} (2s_i-1)
\end{eqnarray*}
where we used the fact that $\log p(\bs)=-g\max(m_\bs-1,0)-\log Z$ and that that $\sum_{\bs}\phi^\mu=0$ for all $\mu>0$. 
The sum in the last line vanishes if the smallest index $i_\mu=\min\{i:i\in\mu\}$ in the sum is less than $k$. Hence
\[
\sum_{\bs\in \mathcal{S}_k} \prod_{i\in\mu} (2s_i-1)=2^{k-1}(-1)^{|\mu|}\left[I_{i_\mu>k}-\delta_{i_\mu,k}\right]
\]
where $|\mu|$ is the number of variables in operator $\phi^\mu$, i.e. the order of the interaction, and $I_{i>j}=1$ if $i>j$ and $I_{i>j}=0$ otherwise. A straightforward calculation leads to
\begin{equation}
\label{ }
g^\mu=\frac{(-1)^{-|\mu|}}{2^n}\left(2^{i_\mu}-2\right)g.
\end{equation}
Interesting all couplings of operators that contain the first feature vanish, because $i_\mu=1$. This is a consequence of the fact that the first feature is independent of all the others, because 
\[
p(s_2,\ldots,s_n|s_1=1)=p(s_2,\ldots,s_n|s_1=0)=p(s_2,\ldots,s_n)
\]
as can be verified explicitly. 
All other operators $\phi^\mu$ have a non-zero coupling, so there are interactions of arbitrary order in this model. Note that the couplings take only few values, depending on $i_\mu$ and $|\mu|$. 

The couplings $J_\nu$ of the model 
\begin{equation}
\label{ }
p(\bs)=e^{\sum_\nu J_\nu\psi^\nu(\bs)},\qquad \psi^\nu(\bs)=\prod_{i\in\nu}s_i
\end{equation}
expressed directly in terms of operators $\psi^\nu$ which are products of the variables $s_i$ can also be derived. Indeed , notice that 
\[
\prod_{i\in\mu}(2s_i-1)=\sum_{\nu\subseteq \mu}2^{|\nu|}(-1)^{|\mu|-|\nu|}\prod_{i\in\nu}s_i\,,
\]
where the sum on $\nu\subseteq\mu$ indicates a sum over all subsets of $\mu$. 
Then the coupling $J_\nu$ is given by the sum of all operators $\mu$ which contain $\nu$ as a subset of indices, i.e. 
\begin{equation}
\label{Jnu}
J_\nu=(-1)^{|\nu|}g 2^{|\nu|-n}
\sum_{\mu:\,\nu\subseteq\mu} \left(2^{i_\mu}-{2}\right).
\end{equation}
This shows that all $J_\nu$ are generally non-zero, apart from those that contain the first feature.
Indeed if $\nu$ contains the first feature (i.e. $i_\nu=1$) then $i_\mu=1$ for all $\mu$ such that $\nu\subseteq\mu$. 
The sum in Eq.~(\ref{Jnu}) can be evaluated splitting it into a sum over $i_\mu\le i_\nu$ and the sum over all operators $\mu$ that contain $\nu$ and whose lowest index variable is $i_\mu$. A careful calculation leads to the surprisingly simple result Eq.~(\ref{Jnu_fin}).

Finally, let us discuss the properties of the HFM under marginalisation of the highest order features. 
In order to do this, it is convenient to specify the dependence of $p(\bs)=p_n(s_1,\ldots,s_n)$ on $n$ and on the variables $s_i$. It is easy to see that the marginal distribution on the highest order feature is given by:
\[
\sum_{s_n=0,1}p_n(s_1,\ldots,s_n)=\frac{Z_{n-1}}{Z_n}p_{n-1}(s_1,\ldots,s_{n-1})+\frac{1}{Z_n} e^{-g(n-1)}\,,
\]
where $Z_n$ is the partition function of the $n$ features HFM.
Iterating this equation $m$ times one finds 
\[
\sum_{s_{n-m+1}=0,1}\cdots \sum_{s_n=0,1}p_n(s_1,\ldots,s_n)=\frac{Z_{n-m}}{Z_n}p_{n-m}(s_1,\ldots,s_{n-m})+
\left(1-\frac{Z_{n-m}}{Z_n}\right)2^{m-n}\,,
\]
This is a mixture between an HFM and a flat distribution. 
For $g>g_c$ the marginal distribution remain close to an HFM.

Finally, let us study the properties of the HFM under the transformation discussed in Section~\ref{sec:universalps}. Consider an HFM with parameter $g$ and $n$ features $\bs=(s_1,\ldots, s_n)$. Each of these features is now duplicated, i.e. 
$s_i\to (\tilde s_{2i-1},\tilde s_{2i})$ where the state $s_i=0$ correspond to the state $\tilde s_{2i-1}=\tilde s_{2i}=0$ and the state $s_i=1$ to the states where at least one of the two features $\tilde s_{2i-1},\tilde s_{2i}$ is set to one.
Assuming that the new state $\tilde\bs$ also follows the HFM distribution with parameter $\tilde g$, we have
\[
P\{\tilde\bs\in\mathcal{S}_{2k-1}\}+P\{\tilde\bs\in\mathcal{S}_{2k}\}=\left[(2e^{-\tilde g})^2\right]^{k-1}(2e^{-\tilde g}+1)\,.
\]
Setting this equal to $P\{\bs\in\mathcal{S}_{k}\}$ yields a transformation $g\to\tilde g$ that for large $k$ reads
\begin{equation}
\label{RG}
\tilde g=\frac 1 2 (\log2+g).
\end{equation}
This transformation has an attractive fixed point at $\tilde g=g=\log 2=g_c$. This means that under a transformation where features are duplicated, the HFM converges to the critical point $g=g_c$. In loose words, the phase space of the HFM 
can be expanded by duplicating features without affecting the relative population of distinct features. The inverse of this transformation can be used to define a renormalisation group transformation where $n$ features $(\tilde s_{n+1},\ldots,\tilde s_{2n})$ are added to an HFM with $n$ features $(\tilde s_1,\ldots,\tilde s_n)$ with parameter $\tilde g$. This leads to an HFM with $2n$ features $\tilde \bs$ and parameter $\tilde g$.
Consider the inverse transformation to the one discussed above, where two consecutive features $(\tilde s_{2i-1},\tilde s_{2i})$ are mapped to one feature $s_i$. Under this coarse graining step, 
it is possible to realise a renormalisation group transformation $(\tilde s_1,\ldots,\tilde s_n)\to (s_1,\ldots, s_n)$ in which the parameter $\tilde g$ is transformed to $g$. This transformation is exactly the inverse of the transformation (\ref{RG}). Hence $g=g_c$ is a repulsive fixed point, whereas $g=0$ and $g=\infty$ are attractive fixed points.

\section{Expansion of the log-likelihood around the featureless state}
\label{app:w1}

In this Section we consider the expansion around the state where $a_j$ is given by Eq.~(\ref{w0sol}) and $w_{i,j}=0$ for all $i,j$. For convenience, we introduce the notation $\sigma_j\equiv\sigma(a_j)=(1+e^{-a_j})^{-1}$ which coincides with the value of $\bar x_j$ in this state. A tedious calculation reveals that
\begin{eqnarray*}
{\Lik} & \simeq & \sum_j \left[a_j\bar x_j +\log(1-\sigma_j)+\sum_i\E{s_i}w_{i,j}(\bar x_j-\sigma_j)\right]\\
 & ~ & +\frac 1 2 \sum_{j,j'}\overline{(x_j-\sigma_j)(x_{j'}-\sigma_{j'})}\sum_{i,i'}
 \left[\E{s_is_{i'}}-\E{s_i}\E{s_{i'}}\right]w_{i,j}w_{i',j'} \\
 & ~ &-\frac 1 2 \sum_j \sigma_j(1-\sigma_j)\sum_{i,i'}\E{s_is_{i'}}
 w_{i,j}w_{i',j} \\
& ~ & + \frac 1 2 \sum_{j,j'} \left[\overline{(x_{j'}-\sigma_{j'})^2}-\sigma_{j'}(1-\sigma_{j'}) \right](\bar x_j-\sigma_j)\sum_{i,i',i''}w_{i,j'}w_{i',j'}w_{i'',j}\left[\E{s_is_{i'}s_{i''}}-\E{s_{i''}}\E{s_is_{i'}}\right]\\
 & ~ &+\frac {1}{2}\sum_j(\bar x_j^3+11\sigma_j^2-10\sigma_j^3-5\sigma_j\bar x_j+8\sigma_j^2\bar x_j-3\sigma_j\bar x_j^2-2\sigma_j)\sum_{i,i',i''}w_{i,j}w_{i',j}w_{i'',j}\E{s_is_{i'}s_{i''}}\\
 & ~ &+\sum_{j<j'<j''}\overline{(x_j-\sigma_j)(x_{j'}-\sigma_{j'})(x_{j''}-\sigma_{j''})} \sum_{i,i',i''}w_{i,j}w_{i',j'}w_{i'',j''}\E{s_is_{i'}s_{i''}}\\
  & ~ &-\sum_{j<j',j''}\overline{(x_j-\sigma_j)(x_{j'}-\sigma_{j'})(x_{j''}-\sigma_{j''})}\sum_{i,i',i''}w_{i,j}w_{i',j'}w_{i'',j''}\E{s_is_{i'}}\E{s_{i''}}\\
  & ~ &+\frac {1}{ 3}\sum_{j,j',j''}\overline{(x_j-\sigma_j)(x_{j'}-\sigma_{j'})(x_{j''}-\sigma_{j''})}\sum_{i,i',i''}w_{i,j}w_{i',j'}w_{i'',j''}\E{s_i}\E{s_{i'}}\E{s_{i''}}\\
\end{eqnarray*}
%


Considering this expansion to second order, one finds that the featureless state is a saddle point. 

\subsection{Adding the first feature}

We now consider the state where $w_{i,j}=0$ for $i>1$ and only the first feature is non-zero. For convenience we set $w_{1,j}=w_j$ and $s_1=s$, and consider the expansion up to quadratic terms. We need to expand also with respect to $a_j$ around the solution Eq.~(\ref{w0sol}). This means that $\sigma_j=\bar x_j+\bar x_j(1-\bar x_j)\delta a_j+\frac 1 2 \bar x_j(1-\bar x_j)(1-2\bar x_j)\delta a_j^2+\ldots$. Inserting this in the previous expression we find 
\begin{eqnarray}
\label{expw1}
{\Lik} & \simeq & \sum_j \left[\bar x_j\log\bar x_j+(1-\bar x_j)\log(1-\bar x_j)\right]\\
& ~&- \sum_j \left[\frac 1 2 \bar x_j(1-\bar x_j)\delta a_j^2+
\E{s}w_{j}\bar x_j(1-\bar x_j)\delta a_j\right]\\
\nonumber
& & +\frac 1 2 
\sum_{j,j'}w_{j}D_{j,j'}w_{j'} + O(w^3)\\
\nonumber
D_{j,j'} & = & \E{s}(1-\E{s})\overline{(x_j-\bar x_j)(x_{j'}-\bar x_{j'})}-\E{s}\bar x_j(1-\bar x_j)\delta_{j,j'}
\end{eqnarray}
The maximisation over $\delta a_j$ is done straightforwardly and 
\[
\delta a_j=-\E{s}w_j
\]
inserting this into Eq.~(\ref{expw1}) leads to Eq.~(\ref{expw1text}) in the main text.

\subsection{Adding the next features}

Let us assume that $a_j$ and $w_{i,j}$ for $i<n$ are fixed whereas $w_{i,j}=0$ for $i\ge n$. The gradients of the log-likelihood read 
\begin{eqnarray}
\frac{\partial {\Lik}}{\partial a_j} & = & \bar x_j-\overline{\E{\sigma(h_j(\bs))|\bx}} =0\\
\frac{\partial {\Lik}}{\partial w_{i,j}} & = & \overline{x_j\E{s_i|\bx}}-\overline{\E{s_i\sigma(h_j(\bs))|\bx}} =0,\qquad i<n
\end{eqnarray}
where $h_j(\bs)=a_j+\sum_{i<n} s_iw_{i,j}$ and $\E{\ldots|\bx}$ stand for expected values over $p(\bs|\bx)$.
Notice that $s_n$ is independent of $\bx$ as long as $w_{n,j}=0$. 
Let us consider the equations for $w_{n,j}$:
\begin{eqnarray}
\frac{\partial {\Lik}}{\partial w_{n,j}} & = & \bar x_j\E{s_n}-\overline{\E{s_n\sigma(h_j(\bs))|\bx}} \\
 & = & \overline{\E{\sigma(h_j(\bs))|\bx}}\E{s_n}-\overline{\E{s_n\sigma(h_j(\bs))|\bx}} \,.
 \label{gradfeat}
\end{eqnarray}
For $n=1$, $h_j(\bs)=a_j$ is independent of $\bs$ and the gradients vanish, as discussed above. For $n=2$ the gradients with respect to $w_{2,j}$ also vanish because $\sigma(h_j(\bs))$,  when $w_{i,j}=0$ for $i>1$,  only depends on $s_1$ which is independent of $s_2$. 

Hence learning has to start from a non-zero value of $w_{i,j}$ for $i\le 2$. 
When $n>2$ the gradients in Eq.~(\ref{gradfeat}) are non-zero even when $w_{n,j}=0$, because $s_n$ is not independent of $h_j(\bs)$ and hence the covariance in the last expression does not vanish.

\section{Details on the numerical experiments}
\label{app:algo}

The results of Section 3.1, 3.2 and 3.3 were obtained with the following algorithm, as described in the main text:

\bigskip
\begin{tabular}{l}
\hline \textbf{Algorithm} Learning one feature at the time\\
\hline
\textbf{Input}: Training dataset $\hat\mathbf{x}$\\
~~~~~~~~~~~Assignment $\hat\mathbf{s}$ with $n\ge 0$ features\\

\textbf{Initialize}: The $n+1^{\rm st}$ hidden feature is turned on ($s^{(a)}=1$) for the $N\xi^n/Z$\\
~~~~~~~~~~~ data points with the smallest likelihood.\\

\textbf{E-Step}: Given $\theta$, find the best assignment: $\hat\bs\leftarrow {\rm arg}\max_{\hat \bs} \tilde\mathcal{L}$. \\

\textbf{M-Step}: Given $\hat\bs$, adjust the parameters: $\theta\leftarrow {\rm arg}\max_{\theta} \tilde\mathcal{L}$\\

\textbf{Repeat} E-step and M-step until convergence. \\

\textbf{Iterate:} $n\leftarrow n+1$, go to Initialize.\\
\hline
\end{tabular}
\bigskip

For the results in Section~\ref{sec:universalps}, training is performed on NVIdia V100 GPU and use 60000 training samples of each dataset, with the initial parameters $\hat{w}$ and $\mathbf{a}$ drawn from a Gaussian distribution with zero mean and variance $0.01$.

We explored different algorithms. As discussed in the main text, these are based on an expectation-maximisation strategy. Given that $p(\bs)$ is fixed, we first draw a sample $\hat\bs=(\bs^{(1)},\ldots,\bs^{(N)})$ of $N$ internal states. Then we alternate M-steps based on the usual gradient ascent with E-steps in which internal states $\bs^{(i)}$ are assigned to data-points $\bx^{(j)}$ in order to maximise the joint log-likelihood. In practice, an assignment coincides with a permutation $j=\pi_i$ of the integers $(1,\ldots,N)$. 

\bigskip
\begin{tabular}{l}
\hline \textbf{Algorithm} Learning all features at once\\
\hline
\textbf{Input}: Training dataset $\hat\mathbf{x}$, the state of hidden units $\hat\mathbf{s}$ is drawn from  $p(\mathbf{s})$, \\ ~~~~~~~~~~~independently for each datapoint. \\

\textbf{Initialize}: $\theta^{(0)}=\{\mathbf{a}, \hat\mathbf{w}\}$ are drawn at random. \\

\textbf{E-Step}: Given $\theta$, find the best assignment: $\hat\bs\leftarrow {\rm arg}\max_{\hat \bs} \tilde\mathcal{L}$. \\

\textbf{M-Step}: Given $\hat\bs$, adjust the parameters: $\theta\leftarrow {\rm arg}\max_{\theta} \tilde\mathcal{L}$\\

\textbf{Repeat} E-step and M-step until convergence. \\
\hline
\end{tabular}

\bigskip
In Step 1 learning is performed in mini-batches of $20$ data points and the learning rate is $0.01$. 
In Step 2, we locate the best assignment of hidden states to the training dataset by using the Jonker-Volgenant algorithm \cite{jonker1987shortest}.

\section{Restricted Boltzmann Machines (RMBs)}
\label{app:RBM}

In a Restricted Boltzmann Machine (RBM) the joint probability distribution $p(\bx,\bs|\theta)$ is given by
\begin{equation}
\label{RBM}
p(\bx,\bs|\theta)=\frac{1}{Z(\theta)} e^{-E(\bx,\bs|\theta)}
\end{equation}
where $Z$ is a normalisation constant and the "energy" function
\begin{equation}
\label{RBME}
E(\bx,\bs|\theta)=\sum_{j=1}^m b_j x_j+\sum_{i=1}^n c_i s_i+\sum_{i,j} s_i u_{i,j} x_j
\end{equation}
depends on the parameters $\theta=\{b_j,\,c_i,\,u_{i,j},~i=1,\ldots,n;~j=1,\ldots,m\}$.

Given a dataset $\hat \bx=(\bx^{(1)},\ldots,\bx^{(N)})$ of $N$ samples of the visible variables, the network is trained by maximising the log-likelihood
\begin{equation}
\label{ }
\mathcal{L}(\hat\bx|\theta)=\sum_{i=1}^N \log p(\bx^{(i)}|\theta),
\end{equation}
over $\theta$, where
\begin{equation}
\label{ }
p(\bx|\theta)=\sum_{\bs} p(\bx,\bs|\theta)
\end{equation}
is the marginal distribution of $\bx$. Training maps the dataset $\hat \bx$ onto an internal representation 
\begin{equation}
\label{eq:ps}
p(\bs|\hat\theta)=\sum_{\bx} p(\bx,\bs|\hat\theta)
\end{equation}
which is the marginal distribution over hidden states $\bs$. Here $\hat \theta={\rm arg}\max_\theta \mathcal{L}(\hat\bx|\theta)$ are the learned parameters after training. We refer to Ref.~\cite{hinton2012practical} for a classical reference to training RBMs and to~\cite{fischer2015training} for a more recent one.

\end{document}